\documentclass[12pt]{article}
\usepackage{amsmath,bbm,amssymb,amsthm,amscd}
\usepackage{epsfig,graphics,longtable}
\usepackage{upgreek}

\theoremstyle{theorem}
\newtheorem{Theorem}{Theorem}[section]
\newtheorem{Proposition}[Theorem]{Proposition}
\newtheorem{Lemma}[Theorem]{Lemma}
\newtheorem{Corollary}[Theorem]{Corollary}
\newtheorem{Definition}[Theorem]{Definition}

\theoremstyle{remark}
\newtheorem{Remark}[Theorem]{Remark}
\newtheorem{Example}[Theorem]{Example}

\newcommand{\btm}{\begin{Theorem}}
\newcommand{\etm}{\end{Theorem}}

\newcommand{\ben}{\begin{enumerate}}
\newcommand{\een}{\end{enumerate}}

\newcommand{\bit}{\begin{itemize}}
\newcommand{\eit}{\end{itemize}}

\newcommand{\bre}{\begin{Remark}\rm}
\newcommand{\ere}{\end{Remark}}

\newcommand*{\bbm}{\begin{Remark}}
\newcommand*{\ebm}{\end{Remark}}

\newcommand{\ble}{\begin{Lemma}}
\newcommand{\ele}{\end{Lemma}}

\newcommand*{\bsz}{\begin{Proposition}}
\newcommand*{\esz}{\end{Proposition}}

\newcommand{\beq}{\begin{equation}}
\newcommand{\eeq}{\end{equation}}

\newcommand{\bbma}{\begin{bmatrix}}
\newcommand{\ebma}{\end{bmatrix}}

\newcommand*{\bbs}{\begin{Example}}
\newcommand*{\ebs}{\end{Example}}

\newcommand*{\bfg}{\begin{Corollary}}
\newcommand*{\efg}{\end{Corollary}}

\newcommand*{\bdf}{\begin{Definition}}
\newcommand*{\edf}{\end{Definition}}

\newcommand*{\bbw}{\begin{proof}}
\newcommand*{\ebw}{\end{proof}}

\newcommand{\CC}{{\mathbb{C}}}

\newcommand{\HH}{{\mathbb{H}}}

\newcommand{\RR}{{\mathbb{R}}}

\newcommand{\ZZ}{{\mathbb{Z}}}
\newcommand{\momap}{\mu}
\newcommand{\rcfg}{{\mc X}}
\newcommand{\rpha}{{\mc P}}
\newcommand{\OT}{{\mc T}}
\newcommand{\RT}{{\tt R}}
\newcommand{\RTT}{{\tt R^\ast}}
\newcommand{\rt}{{\tt r}}
\newcommand{\rtt}{{(\rt)}}
\newcommand{\pr}{\mr{pr}}
\newcommand{\hw}{\lambda}
\newcommand{\HW}{\Lambda_+}
\newcommand{\kf}{k}
\newcommand{\RS}{\Sigma}
\newcommand{\RSS}{\Gamma}
\newcommand{\SSR}{\Pi}
\newcommand{\PS}{\Pi}
\newcommand{\CSS}{\Delta}    
\newcommand{\WS}{G} 
\newcommand{\HWS}{G_+} 

\newcommand{\rel}[1]{\nu_{#1}}
\newcommand{\efn}[1]{E_{#1}}
\newcommand{\vfh}[1]{V_{#1}}
\newcommand{\tvfh}[1]{\tilde V_{#1}}
\newcommand{\vfhp}[1]{V^+_{#1}}
\newcommand{\vfhm}[1]{V^-_{#1}}
\newcommand{\ncc}{\tilde\chi^\CC}  

\newcommand{\rank}{\mr{rank}}
\newcommand{\sign}{\mr{sign}}

\renewcommand{\sl}{\mf{sl}}
\newcommand{\SO}{{\mr{SO}}}

\newcommand{\Sp}{{\mr{Sp}}}
\renewcommand{\sp}{{\mf{sp}}}
\newcommand{\Spin}{{\mr{Spin}}}

\newcommand{\SU}{{\mr{SU}}}

\newcommand{\minus}{\hspace{-0.1pt}\scriptscriptstyle-\hspace{-0.1pt}}
\newcommand{\plus}{\hspace{-0.25pt}\text{\tiny$\scriptscriptstyle+$}\hspace{-0.25pt}}

\newcommand{\al}[1]{\begin{align} #1 \end{align}}
\newcommand{\ala}[1]{\begin{align*} #1 \end{align*}}

\DeclareMathOperator{\Ad}{Ad}
\DeclareMathOperator{\diag}{diag}

\newcommand{\mc}[1]{\mathcal{#1}}
\newcommand{\mf}[1]{\mathfrak{#1}}
\newcommand{\mr}[1]{\mathrm{#1}}

\newcommand{\Hi}{{\mc H}}

\newcommand{\comment}[1]{}
\newcommand{\verweis}[1]{}
\newcommand{\todo}[1]{}
\renewcommand{\d}{{\mr d}}

\newcommand{\cfg}{{\mc X}}
\newcommand{\pha}{{\mc P}}

\newcommand{\vp}{\varphi}

\newcommand{\ctg}{\mr T^\ast}

\newcommand{\nm}{\mr N}
\newcommand{\rref}[1]{{\rm \ref{#1}}}

\newcommand{\ol}[1]{\overline{#1}}
\newcommand{\abs}{\hspace*{2.5mm}}
\newcommand*{\qeb}{\nopagebreak\hspace*{0.1em}\hspace*{\fill}{\mbox{\small$\blacklozenge$}}}

\newcommand{\linie}[3]{\put(#1){\line(#2){#3}}}
\newcommand{\marke}[3]{\put(#1){\makebox(0,0)[#2]{$\scriptstyle#3$}}}
\newcommand{\punkt}[1]{\put(#1){\circle*{0.175}}}

\newcommand{\kreis}[1]{\put(#1){\circle{0.175}}}
\newcommand{\kkreis}[1]{\put(#1){\circle{0.25}}}

\newcommand{\lr}[1]{\put(#1){\linie{0.1,0}{1,0}{0.8}}}
\newcommand{\lo}[1]{\put(#1){\linie{0,0.1}{0,1}{0.8}}}
\newcommand{\lro}[1]{\put(#1){\linie{0.0707,0.0707}{1,1}{0.8586}}}
\newcommand{\lru}[1]{\put(#1){\linie{0.0707,-0.0707}{1,-1}{0.8586}}}

\newcommand{\hlr}[1]{\put(#1){\linie{0.1,0}{1,0}{0.4}}}
\newcommand{\hll}[1]{\put(#1){\linie{-0.1,0}{-1,0}{0.4}}}
\newcommand{\hlo}[1]{\put(#1){\linie{0,0.1}{0,1}{0.4}}}
\newcommand{\hlu}[1]{\put(#1){\linie{0,-0.1}{0,-1}{0.4}}}

\newcommand{\dvlro}[1]{\put(#1){\linie{0.0707,0.0707}{1,1}{0.6439}}}
\newcommand{\dvllo}[1]{\put(#1){\linie{-0.0707,0.0707}{-1,1}{0.6439}}}
\newcommand{\dvlru}[1]{\put(#1){\linie{0.0707,-0.0707}{1,-1}{0.6439}}}
\newcommand{\dvllu}[1]{\put(#1){\linie{-0.0707,-0.0707}{-1,-1}{0.6439}}}
\newcommand{\vlro}[1]{\put(#1){\linie{0.0707,0.0707}{1,1}{0.285}}}
\newcommand{\vllo}[1]{\put(#1){\linie{-0.0707,0.0707}{-1,1}{0.285}}}
\newcommand{\vlru}[1]{\put(#1){\linie{0.0707,-0.0707}{1,-1}{0.285}}}
\newcommand{\vllu}[1]{\put(#1){\linie{-0.0707,-0.0707}{-1,-1}{0.285}}}

\newcommand{\Ateil}[1]{\put(#1){
 \put(0,0){\circle{0.2}} 
 \put(0.1,0){\line(1,0){0.4}}
 \put(0.75,0){\circle*{0.01}}
 \put(1,0){\circle*{0.01}}
 \put(1.25,0){\circle*{0.01}}
 \put(1.5,0){\line(1,0){0.4}}
 \put(2,0){\circle{0.2}}
}}

\newcommand{\ACteil}[1]{\put(#1){
 \put(0,0){\circle*{0.2}} 
 \put(0.1,0){\line(1,0){0.4}}
 \put(0.75,0){\circle*{0.01}}
 \put(1,0){\circle*{0.01}}
 \put(1.25,0){\circle*{0.01}}
 \put(1.5,0){\line(1,0){0.4}}
 \put(2,0){\circle*{0.2}}
}}

\newcommand{\Bre}[1]{\put(#1){\put(0,0){\put(0,0){\circle*{0.2}}
\multiput(0,0.095)(0,-0.19){2}{\line(1,0){1}}}}}
\newcommand{\Bli}[1]{\put(#1){\put(0,0){\put(0,0){\circle*{0.2}}
\multiput(0,0.095)(0,-0.19){2}{\line(-1,0){1}}}}}
\newcommand{\Cre}[1]{\put(#1){\put(0,0){\put(0,0){\circle{0.2}}
\multiput(0,0.095)(0,-0.19){2}{\line(1,0){1}}}}}
\newcommand{\Cli}[1]{\put(#1){\put(0,0){\put(0,0){\circle{0.2}}
\multiput(0,0.095)(0,-0.19){2}{\line(-1,0){1}}}}}
\newcommand{\Dre}[1]{\put(#1){\multiput(0,0.4)(0,-0.8){2}{\circle{0.2}}\put(0.907,0.037){\line(-5,2){0.81}}\put(0.907,-0.037){\line(-5,-2){0.81}}}}
\newcommand{\DCre}[1]{\put(#1){\multiput(0,0.4)(0,-0.8){2}{\circle*{0.2}}\put(0.907,0.037){\line(-5,2){0.81}}\put(0.907,-0.037){\line(-5,-2){0.81}}}}
\newcommand{\Dli}[1]{\put(#1){\multiput(0,0.4)(0,-0.8){2}{\circle{0.2}}\put(-0.907,0.037){\line(5,2){0.81}}\put(-0.907,-0.037){\line(5,-2){0.81}}}}
\newcommand{\Pure}[1]{\put(#1){\multiput(0.15,0)(0.35,0){3}{\circle*{0.01}}}}

\newcommand{\hpunkt}[1]{\put(#1){\circle*{0.075}}}
\newcommand{\opunkt}[1]{\put(#1){\circle{0.075}}}

\newcommand{\whole}[3]{\put(#1){\hpunkt{0,0}\put(0.1,0.1){\makebox(-0.2,
   -0.2)[cl]{\tiny \fbox{$#3$}}}}}

\newcommand{\lri}[3]{\put(#1){\hpunkt{0,0}\linie{0.1,0}{1,0}{0.8}
\put(0.0375,0.075){\makebox(-0.075,-0.15)[#2]{\tiny $#3$}}}}
\newcommand{\lrii}[3]{\put(#1){\hpunkt{0,0}\linie{0.1,0}{1,0}{1.8}
\put(0.0375,0.075){\makebox(-0.075,-0.15)[#2]{\tiny $#3$}}}}

\newcommand{\lori}[3]{\put(#1){\hpunkt{0,0}\linie{0.0894,0.0447}{2,1}{0.821}
\put(0.0375,0.075){\makebox(-0.075,-0.15)[#2]{\tiny $#3$}}}}

\newcommand{\luri}[3]{\put(#1){\hpunkt{0,0}\linie{0.0894,-0.0447}{2,-1}{
0.821}\put(0.0375,0.075){\makebox(-0.075,-0.15)[#2]{\tiny $#3$}}}}
\newcommand{\lurii}[3]{\put(#1){\hpunkt{0,0}\linie{0.0894,-0.0447}{2,-1}{
1.821}\put(0.0375,0.075){\makebox(-0.075,-0.15)[#2]{\tiny $#3$}}}}

\newcommand{\lorri}[3]{\put(#1){\hpunkt{0,0}\linie{0.0968,0.0242}{4,1}{1.8064}
\put(0.0375,0.075){\makebox(-0.075,-0.15)[#2]{\tiny $#3$}}}}

\newcommand{\lurri}[3]{\put(#1){\hpunkt{0,0}\linie{0.0968,-0.0242}{4,-1}{
1.8064}\put(0.0375,0.075){\makebox(-0.075,-0.15)[#2]{\tiny $#3$}}}}

\newcommand{\olri}[3]{\put(#1){\opunkt{0,0}\linie{0.1,0}{1,0}{0.8}
\put(0.0375,0.075){\makebox(-0.075,-0.15)[#2]{\tiny $#3$}}}}

\newcommand{\olori}[3]{\put(#1){\opunkt{0,0}\linie{0.0894,0.0447}{2,1}{0.821}
\put(0.0375,0.075){\makebox(-0.075,-0.15)[#2]{\tiny $#3$}}}}

\newcommand{\oluri}[3]{\put(#1){\opunkt{0,0}\linie{0.0894,-0.0447}{2,-1}{
0.821}\put(0.0375,0.075){\makebox(-0.075,-0.15)[#2]{\tiny $#3$}}}}
\newcommand{\olurii}[3]{\put(#1){\opunkt{0,0}\linie{0.0894,-0.0447}{2,-1}{
1.821}\put(0.0375,0.075){\makebox(-0.075,-0.15)[#2]{\tiny $#3$}}}}

\begin{document}

\title{\bf On the Reflection Type Decomposition of the Adjoint Reduced Phase Space of a Compact Semisimple Lie group}

\author{
M.\ Hofmann$^\dagger$, G.\ Rudolph$^\ast$ and M.\ Schmidt$^\ast$
\\[5pt]
$^\dagger$ Naturwissenschaftlich-Technische Fakult\"at, Universit\"at Siegen 
\\
Walter-Flex-Str.\ 3, 57068 Siegen, Germany
\\[5pt]
$^\ast$ Institut f\"ur Theoretische Physik, Universit\"at Leipzig
\\
Augustusplatz 10/11, 04109 Leipzig, Germany
\\
    }

\maketitle

\comment{
\vspace{2cm}

{\bf Keywords:}~
\\

}
\vspace{2cm}

\begin{abstract}

\noindent
We consider a system with symmetries whose configuration space is a compact Lie
group, acted upon by inner automorphisms. The classical reduced phase space of
this system decomposes into connected components of orbit type subsets. To
investigate hypothetical quantum effects of this decomposition one has to
construct the associated costratification of the Hilbert space of the quantum system in
the sense of Huebsch\-mann. In the present paper, instead of the decomposition
by orbit types, we consider the related decomposition by reflection types
(conjugacy classes of reflection subgroups). These two decompositions turn out
to coincide e.g.\ for the classical groups $\SU(n)$ and $\Sp(n)$. We derive
defining relations for reflection type subsets in terms of irreducible characters 
and discuss how to obtain from that the corresponding co\-strat\-ification of
the Hilbert space of the system. To illustrate the method, we give explicit
results for some low rank classical groups.  

\end{abstract}

\newpage


\section{Introduction}
\label{S-intro}


This paper is part of a program which aims at developing a non-perturbative
approach to the quantum theory of gauge fields in the Hamiltonian framework 
with special emphasis on the role of non-generic gauge orbit strata. 
The starting point is a finite-dimensional lattice approximation of the
theory.\footnote{Of course, ultimately, one is interested in constructing the
continuum limit, see \cite{GR} for a first step in this direction.}  In this
context, one obtains   
a finite-dimensional Hamiltonian system with symmetries and one can perform 
Marsden-Weinstein reduction. The configuration space and, consequently, the phase space of 
this system both have a stratified structure, which was studied in \cite{cfg, cfgtop,FRS}. 
On quantum level, we have investigated the structure of the observable algebra
\cite{qcd1, qcd2, qcd3, RS}, including the study of the superselection
structure, and we have obtained some insight into the the role of the
non-generic gauge orbit strata \cite{kaehler,HRS,RSV}.
The latter is based on an idea of Huebschmann and will be explained in Section \ref{A-adjQuo} in detail. In short, it may be 
described as follows. One views the states of the unreduced system as elements of the Hilbert space $\cal H$ of 
square integrable holomorphic functions, implements the symmetry reduction on quantum level and, for a given classical gauge orbit stratum, one considers 
the subspace of $\cal H$ consisting of functions vanishing on this stratum. Next, one  
takes its orthogonal complement in $\cal H$. This is, by definition, the Hilbert space of physical 
states associated with the stratum under consideration. Performing this procedure for every 
element of the stratification, one obtains a partially ordered family of closed subspaces of $\cal H$ called the costratification associated with the classical gauge orbit stratification. In \cite{HRS} we studied this procedure for a toy model with gauge group $\SU(2)$. 

The aim of the present paper is to make a step towards extending the results of \cite{HRS} to an arbitrary compact connected semisimple Lie group $G$. 
Let us explain the setting. We consider a physical system with configuration space $G$, acted upon by itself by inner automorphisms. As usual, we will refer to this action as the adjoint action of $G$. This models a lattice gauge theory with structure group $G$ on a single plaquette in a tree gauge, together with the residual gauge transformations. The phase space is given by the cotangent bundle $\ctg G$, acted upon by the lifted action. This action is symplectic and possesses a natural momentum mapping $\momap : \ctg G \to \mf g^\ast$, where $\mf g$ denotes the Lie algebra of $G$ \cite{AbrahamMarsden}. Using an invariant scalar product $\langle \cdot,\cdot
\rangle$ on $\mf g$ and left translation on $G$, we can define a global
trivialization of $\ctg G$ by  
\beq\label{G-Trvis}
G\times\mf g \to \ctg G
 \,,~~~~~~
(a,A) \mapsto \langle A_a,\cdot \rangle\,,
\eeq
where $A$ has to be interpreted as a left-invariant vector field and $A_a$ means the value of $A$ at $a$. In this trivialization, the lifted action reads
$$
g\cdot(a,A) = \big(gag^{-1},\Ad(g)A\big)\,,
$$
where $a,g\in G$ and $A\in\mf g$, and the momentum mapping is given by 
\beq\label{G-ImpAbb}
\momap(a,A) = \Ad(a) A - A\,.
\eeq

The reduced configuration space $\rcfg$ is given by the adjoint quotient, that
is, by the set of orbits of the adjoint action, endowed with the quotient
topology. Let $\OT$ denote the set of orbit types of the adjoint action. For
$\tau\in\OT$, let $\rcfg_\tau$ denote the subset of $\rcfg$ consisting of the
orbits of type $\tau$. A general theorem on proper Lie group actions implies
that the family $\{\rcfg_{\tau,i}\}$, where $\tau\in\OT$ and $i$ labels the
connected components of $\cfg_\tau$, forms a stratification of $\rcfg$
\cite{Pflaum}. This means, in particular, that the connected
components $\rcfg_{\tau,i}$ are smooth manifold and that the condition of the
frontier holds: if $\rcfg_{\tau_2,i_2}$ intersects the closure
$\ol{\rcfg_{\tau_1,i_1}}$ of $\rcfg_{\tau_1,i_1}$ in $\rcfg$, then
$\rcfg_{\tau_2,i_2} \subset \ol{\rcfg_{\tau_1,i_1}}$. The condition of the
frontier gives rise to a natural partial ordering on the family
$\{\rcfg_{\tau,i}\}$: $\rcfg_{\tau_1,i_1} \leq \rcfg_{\tau_2,i_2}$
iff $\rcfg_{\tau_2,i_2} \subset \ol{\rcfg_{\tau_1,i_1}}$. One can show that this
partial ordering is compatible with the natural partial ordering of orbit types
by inclusion modulo conjugacy. That is, 
\beq\label{G-OT-HO}
\rcfg_{\tau_1,i_1} \leq \rcfg_{\tau_2,i_2}
 ~~~\Rightarrow~~~
\tau_1 \leq \tau_2\,,
\eeq
where $\tau_1 \leq \tau_2$ if there exist representatives $H_1$ of $\tau_1$ and $H_2$ of $\tau_2$ such that $H_1\subseteq H_2$.

The reduced phase space $\rpha$ is obtained from $\ctg G$ by singular symplectic
reduction at zero level. That is, $\rpha$ is the set of orbits of the lifted
action of $G$ on the invariant subset $\momap^{-1}(0) \subseteq \ctg G$, endowed
with the quotient topology induced from the relative topology on this subset. In
lattice gauge theory, the condition $\mu=0$ corresponds to the Gau{\ss} law
constraint. As is true in general for the lift of a Lie group action and the
associated momentum mapping, the action on $\momap^{-1}(0)$ has the same orbit
types as the original action, i.e., $\OT$. By the procedure of singular
symplectic reduction, the connected components $\pha_{\tau,i}$ of the orbit type
subsets $\rpha_\tau$ of $\rpha$, $\tau\in\OT$, are endowed with natural
symplectic manifold structures. Moreover, they provide a stratification of
$\pha$ \cite{SjamaarLerman,OrtegaRatiu}. In particular,
the condition of the frontier holds for the subsets $\rpha_{\tau,i}$ as well and
the induced partial ordering on the family $\{\rpha_{\tau,i}\}$ is compatible
with the natural partial ordering of $\OT$ in the sense of \eqref{G-OT-HO}. The
bundle projection $\pi : \ctg G \to G$ induces a mapping $\hat\pi : \rpha \to
\rcfg$. This mapping is surjective, because $\momap$ is linear on the fibres of
$\ctg G$ and hence $\momap^{-1}(0)$ contains the zero section of $\ctg G$. Note
that $\hat\pi$ need not preserve the orbit type, but, as $\pi$ is equivariant,
one has at least 
$$
\hat\pi(\rpha_\tau) \subseteq \bigcup_{\tau'\geq\tau} \rcfg_{\tau'}\,.
$$


\section{Stratified quantum theory on an adjoint quotient}
\label{A-adjQuo}


To investigate hypothetic quantum effects of the orbit type decomposition of the
reduced phase space, one has to implement this decomposition on quantum level.
Following Huebschmann \cite{kaehler,HRS}, this can be done by means of a family
of closed subspaces of the Hilbert space as follows. First, we pass to the
quantum theory of the reduced system by means of geometric (K\"ahler)
quantization on $\ctg G$ and subsequent reduction. Let $\mf g^\CC$ denote the
complexification of $\mf g$ and let $G^\CC$ denote the complexification of $G$
\todo{siehe Hall wg genauer Definition}. 
The inverse of the polar decomposition on $G^\CC$ yields a diffeomorphism
$$
G\times\mf g \to G^\CC
 \,,~~~~~~
(a,A) \mapsto a\exp(\mr i A)\,,
$$
where $\exp$ denotes the exponential mapping of $G^\CC$. 
Composition of this diffeomorphism with the global trivialization
\eqref{G-Trvis} yields a diffeomorphism $\ctg G \cong G^\CC$ which intertwines
the lifted action of $G$ on $\ctg G$ with the action of the subgroup $G \subset
G^\CC$ on $G^\CC$ by inner automorphisms. Via this diffeomorphism, the
symplectic structure of $\ctg G$ and the complex structure of $G^\CC$ combine to
a K\"ahler structure. Half-form K\"ahler quantization on
$G^\CC$ yields the Hilbert space $HL^2(G^\CC,\d\nu)$ of holomorphic functions on
$G^\CC$ which are square-integrable with respect to a certain measure, $\d\nu$,
containing the Liouville measure on $\ctg G$, the K\"ahler potential on $G^\CC$
and the half-form correction \cite{Hall:cptype}. Reduction then yields the
closed subspace $\mc H = HL^2(G^\CC,\d\nu)^G$ of $G$-invariants as the Hilbert
space of the reduced system. For convenience, we realize $\mc H$ in two
different ways. First, the holomorphic Peter-Weyl theorem \cite{Hue:HPWT}
implies that $\mc H$ is spanned by the irreducible characters $\chi^\CC_\hw$ of
$G^\CC$. Here, $\hw\in\HW$, where $\HW$ denotes the set of the highest weights
of the finite-dimensional irreducible complex representations of $G^\CC$,
relative to some chosen dominant Weyl chamber in some chosen Cartan subalgebra
of $\mf g^\CC$. The system $\{\chi^\CC_\hw : \hw\in\HW\}$ is  orthogonal but not
normalized. According to \cite{Hall:cptype} or \cite[Lemma 3.3]{Hue:HPWT}, the norms are 
$$
N_\hw := \|\chi_\hw^\CC\|
 = 
(\pi\hbar)^{\dim G/4} \, \mr e^{\hbar \|\lambda + \delta\|^2/2}\,,
$$
where $\hbar$ is the Planck constant, the norm $\|\cdot\|$ on $(\mf g^\CC)^\ast$ is defined by a chosen invariant scalar product on $\mf g$ and $\delta$ denotes half the sum of the roots which are positive in the sense of the chosen dominant Weyl chamber,
\beq\label{G-D-delta}
\delta = \frac 1 2 \sum_{\alpha\in\RS_+} \alpha\,.
\eeq
Both $\hbar$ and the scaling of the invariant scalar product appear as parameters in the
measure $\mr d\nu$. Thus, using the orthonormal basis of normed complex characters 
$$
\ncc_\hw := \frac{\chi^\CC_\hw}{N_\hw} \,,~~~~~~ \hw\in\HW\,,
$$
and endowing $\HW$ with the lexicographic ordering defined by the system of
simple roots associated with the chosen Weyl chamber, we may identify $\mc H$
with the Hilbert space $\ell^2$ over the index set $\HW$. Second, the ordinary
Peter-Weyl theorem for $G$ implies that the irreducible characters $\chi_\hw$ of
$G$ provide an orhonormal basis in the closed subspace $L^2(G,\d\mu)^G$ of
$G$-invariant functions in the Hilbert space $L^2(G,\d\mu)$ of functions on $G$
which are square-integrable with respect to the normalized Haar measure $\d\mu$.
Consequently, the assignment of $\chi_\hw$ to $\chi^\CC_\hw/\|\chi^\CC_\hw\|$
for every $\hw\in\HW$ yields a Hilbert space isomorphism 
\beq\label{G-HL2}
\mc H \cong L^2(G,\d\mu)^G\,.
\eeq

\bbm

The Hilbert space isomorphism \eqref{G-HL2} coincides with the
restriction to the $G$-invariants of the Segal-Bargmann transformation
$L^2(G,\d\mu) \to HL^2(G^\CC,\nu)$. Originally, the Segal-Bargmann
transformation for compact Lie groups was developed via heat kernel analysis on
$G$ and $G^\CC$ in \cite{Hall:SB}. Alternatively, as a consequence of the
holomorphic Peter-Weyl theorem cited above, it can be described in terms of
representative functions on $G$ and $G^\CC$. This is the approach we have used
here.  
\qeb

\ebm

Now, we can define the costratified Hilbert space structure associated with the
decomposition of $\rpha$ by the connected components $\rpha_{\tau,i}$ of the
orbit type subsets of $\rpha$. Since the elements of $\mc H$ are holomorphic
functions on $G^\CC$ (and not classes of functions as in the 
$L^2$-case), and since they are $G$-invariant, they descend to continuous
functions on $\rpha$. Thus, to every stratum $\rpha_{\tau,i}$ there corresponds
a subspace $\mc V_{\tau,i}$ consisting of the functions in $HL^2(G^\CC,\d\nu)^G$
which vanish on $\rpha_{\tau,i}$. Then, the subspace $\mc H_{\tau,i}$ associated
with the stratum $\rpha_{\tau,i}$ is defined to be the orthogonal complement of
$\mc V_{\tau,i}$ in $HL^2(G^\CC,\d\nu)^G$. If $\rpha_{\tau_1,i_1} \leq
\rpha_{\tau_2,i_2}$, then $\rpha_{\tau_2,i_2} \subset \ol{\rpha_{\tau_1,i_1}}$ and 
hence $\mc V_{\tau_1,i_1} \subseteq \mc V_{\tau_2,i_2}$. Then, 
$\mc H_{\tau_2,i_2} \subseteq \mc H_{\tau_1,i_1}$, so that to every pair 
$\rpha_{\tau_1,i_1},\rpha_{\tau_2,i_2}$ satisfying $\rpha_{\tau_1,i_1} \leq
\rpha_{\tau_2,i_2}$ there corresponds a bounded linear mapping
$\Pi_{\tau_1,i_1;\tau_2,i_2} : \mc H_{\tau_1,i_1} \subseteq \mc H_{\tau_2,i_2}$, given 
by orthogonal projection. The families $\{\Hi_{\tau,i}\}$ and
$\{\Pi_{\tau_1,i_1;\tau_2,i_2}\}$ constitute a costratified Hilbert space in the sense of Huebschmann. 

The reader will have noticed that the construction of the subspaces $\mc
V_{\tau,i}$ and $\mc H_{\tau,i}$ does not make use of the fact that the
decomposition of $\pha$ 
under consideration is by orbit type connected components. It can be constructed
for any decomposition fulfilling the condition of the frontier. In fact, in this
paper, we will not consider the decomposition of $\pha$ by orbit type connected
components but the decomposition by reflection types $\rt$, to
be defined below. The reason is that this decomposition can be conveniently
described in terms of roots. In many cases, like $\SU(n)$ and $\Sp(n)$, the two decompositions coincide, 
see Remark \ref{Bem-SU(n)}. In general, the decomposition by reflection types is slightly coarser.

Thus, our goal is to find the subspaces $\mc V_\rt
\subseteq HL^2(G^\CC,\d\nu)$ of functions vanishing on the reflection type
stratum $\rpha_\rt$. We will also make some comments on how to get hands on
their orthogonal complements $\mc H_\rt$. A detailed study of the latter
remains as a future task. To determine $\mc V_\rt$, we use that if
$\nu_1,\dots,\nu_l$ are elements of
$HL^2(G^\CC,\d\nu)$ such that $\rpha_\rt = \bigcap_{i=1}^l \nu_i^{-1}(0)$, then
$\mc V_\rt$ is spanned by the functions  
$$
\nu_k \chi^\CC_\hw
 \,,~~~~~~
k=1,\dots,l,~~~~~~ \hw\in\HW\,.
$$
Thus, for every $\rt$ we have to find a set of relations defining $\rpha_\rt$
as a subset of $\rpha$.


\section{Reduction to the Weyl group action}
\label{A-Grdl}


First, we show that, in the special situation under consideration, $\rpha$ can be identified with the quotient of a Lie group action on a manifold. This is not true in general. 

Let $T$ be some chosen maximal toral subgroup of $G$, let $\mf t$ be the Lie algebra of $T$ and let $W=\mr N(T)/\mr C(T)$ be the corresponding Weyl group. Via the adjoint representation, $W$ acts on $\mf t$. Let $T^\CC$ be the image of $T\times\mf t$ under the inverse of the polar decomposition, that is, $T^\CC = T \exp(\mr i \mf t)$. Since conjugation by elements of $\mr C(T)$ leaves $T^\CC$ pointwise invariant, the action of $W$ on $T$ extends to an action of $W$ on $T^\CC$. The polar decomposition is equivariant with respect to this action. Let $\mf t^\CC$ and $\mf g^\CC$ denote the Lie algebras of $T^\CC$ and $G^\CC$, respectively. Then, $\mf t^\CC$ is a Cartan subalgebra of $\mf g^\CC$. Moreover, $\mf t$ and $\mf g$ can be identified in an obvious way with subsets of $\mf t^\CC$ and $\mf g^\CC$, respectively.

\ble

The natural inclusion mapping $T^\CC \to G^\CC$ descends to a homeomorphism from the topological quotient $T^\CC/W$ onto $\rpha$. 

\ele

In particular, in the present situation, the reduced phase space is an orbifold. The homeomorphism provided by the lemma induces a natural projection 
$$
\pr : T^\CC \to \pha\,.
$$

\bbw

First, we show that for every $x\in G^\CC$ satisfying $\momap(x) = 0$, there
exists $g\in G$ such that $gxg^{-1} \in T^\CC$. Let $(a,A)\in G \times \mf g$
such that $x = a \exp(\mr iA)$. According to \eqref{G-ImpAbb}, $\momap(a,A) = 0$
implies $\Ad(a)A=A$. Then, $\exp(tA)$ belongs to the identity component of the
centralizer of $a$ in $G$ for all $t\in\RR$. By Cor.\ 3 of Thm.\ 2 in \S IX.2.2
of \cite{Bou:Lie}, this component coincides with the union of all maximal tori in
$G$ which contain $a$. Hence, there exists a maximal torus containing $a$ and 
$\exp(tA)$ for all $t$. Since all maximal tori are conjugate in $G$, there
exists $g\in G$ such that both $gag^{-1}$ and $g \exp(tA) g^{-1} = \exp
\big(t(\Ad(g) A\big)$ belong to $T$, for all $t$. It follows that $\Ad(g) A\in
\mf t$ and hence $g \cdot (a,A)\in T\times\mf t$. Then, $gxg^{-1} \in T^\CC$. 

Next, we show that two elements $x,y$ of $T^\CC$ are conjugate under $G$ iff
they are conjugate under $W$. For that purpose, it suffices to show that
conjugacy under $G$ implies conjugacy under the normalizer $\nm_G(T)$ of $T$ in
$G$. Thus, let $g\in G$ such that $gxg^{-1} = y$. Let $(a,A), (b,B)\in T \times
\mf t$ such that $x = a \exp(\mr i A)$ and $y = b \exp(\mr i B)$. Then, $g a
g^{-1} = b$ and $\Ad(g) A = B$, so that under conjugation by $g$, the subset
$\{a,\exp t A\}$ of $T$ is  mapped to the subset $\{b,\exp t B\}$ of $T$ for all
$t$. By Cor.\ 7
of Thm.\ 2 in \S IX.2.2 of \cite{Bou:Lie}, there exists $h\in G$ such that 
$gh T (gh)^{-1} = T$, $gh a (gh)^{-1} = b$ and $gh \exp t A (gh)^{-1} = \exp t
B$. The latter implies $\Ad(gh)A = B$ and hence $gh x (gh)^{-1} = y$. 

As a result of these considerations, the natural inclusion mapping $T\times\mf t \to G \times\mf g$ descends to a bijection from $(T \times \mf t)/W$ onto $\rpha$. The induced mapping is a homeomorphism: continuity follows from the fact that the preimage under the natural inclusion mapping $T\times\mf t \to \mu^{-1}(0)$ of a subset of $\mu^{-1}(0)$ which is saturated with respect to the natural projection $\mu^{-1}(0) \to \pha$ is saturated with respect to the natural projection $T\times \mf t \to (T\times \mf t)/W$. Openness follows from the fact that for a Lie group action, both the action mapping and the natural projection to the orbit space are open mappings.
\ebw

Let $\RS$ denote the root system of $\mf g^\CC$ relative to the Cartan subalgebra $\mf t^\CC$. Since $\mf g^\CC$ is semisimple, the Killing form $\kf$ defines a vector space isomorphism between $\mf t^\CC$ and its dual. For $\alpha\in\Sigma$, let $H_\alpha$ denote the image of $\alpha$ under this isomorphism. That is,
$$
\kf(H_\alpha,X) = \alpha(X)
$$
for all $X\in\mf t^\CC$. The action of $W$ on $T^\CC$ induces a faithful representation of $W$ on $\mf t^\CC$ which is orthogonal with respect to $\kf$ and whose image is generated by the mappings 
\beq\label{G-Refl-1}
\sigma_\alpha : \mf t^\CC \to \mf t^\CC
 \,,~~~~~~
\sigma_\alpha(X) = X - 2\frac{\alpha(X)}{\kf(\alpha,\alpha)} H_\alpha\,,
\eeq
where $\alpha\in\RS$. Geometrically, $\sigma_\alpha$ corresponds to the $\kf$-orthogonal reflection about the hyperplane which is $\kf$-orthogonal to $H_\alpha$. To $\sigma_\alpha$ there corresponds a unique element of $W$ which will be denoted by the same symbol and which will be referred to as the reflection associated with $\alpha$. 
Via the vector space isomorphism $\mf t^\CC \cong (\mf t^\CC)^\ast$ induced by
$\kf$, $\kf$ defines a symmetric bilinear form on $(\mf t^\CC)^\ast$, denoted by the
same symbol, and the representation of $W$ on $\mf t^\CC$ induces a
representation of $W$ on $(\mf t^\CC)^\ast$. Since the root system $\RS \subset
(\mf t^\CC)^\ast$ is invariant under this representation, the latter induces an
action of $W$ on $\RS$. For the reflections $\sigma_\alpha$, \eqref{G-Refl-1}
implies 
\beq\label{G-Refl-2}
\sigma_\alpha : \RS \to \RS
 \,,~~~~~~
\sigma_\alpha(\beta) = \beta - 2\frac{\kf(\alpha,\beta)}{\kf(\alpha,\alpha)} \alpha\,.
\eeq
Moreover, using that $\kf$ is $W$-invariant, from \eqref{G-Refl-2} one can conclude that for any $w\in W$,
\beq\label{G-Weyl-Kjg}
\sigma_{w(\alpha)} = w \circ \sigma_\alpha \circ w^{-1}\,.
\eeq

The real vector space spanned by $H_\alpha$, $\alpha\in\RS$, can be identified
with the real subspace $\mr i \mf t \subset \mf t^\CC$. Correspondingly, the
real vector space spanned by $\RS$ can be identified with the dual of $\mr i \mf
t$. Moreover, the lattice generated by the elements 
$2\pi\mr i \frac{2 H_\alpha}{\kf(\alpha,\alpha)}$ is contained in $\ker(\exp)
\subset \mf t$ and it coincides with the latter if $G$ is simply connected
\cite[\S V.2.16]{BroeckertomDieck}.\footnote{Note that in
\cite{BroeckertomDieck}, the system $R$ of real roots is used. One has $R =
\frac{1}{2\pi\mr i} \RS$.}


\section{Reflection types}
\label{A-RT}


A subgroup of $W$ generated by reflections is commonly referred to as a reflection subgroup. The starting point of our analysis is the well-known fact that the stabilizer of an element $X$ of $\mf t^\CC$ under $W$ is a reflection subgroup. In fact, it is generated by the reflections $\sigma_\alpha$ for all $\alpha\in\RS$ with $\alpha(X)=0$ \cite[\S 4.1]{Hsi}.
This may however not be true for elements of $T^\CC$:

\bbs

Consider the projective unitary group $G = \mr P\mr U(3)$, given by the quotient
of the special unitary group $\SU(3)$ by its center. Let $T$ be the subgroup of
cosets consisting of diagonal matrices. The Weyl group $W$ acts on these cosets
by permuting the matrix entries of some representative. It is easy to see that
the stabilizer of the coset of the matrix  
$$
 \bbma 
1 & 0 \\ 0 & \mr e^{\mr i \frac{2\pi}{3}} & 0 \\ 0 & 0 & \mr e^{\mr i \frac{4\pi}{3}} 
 \ebma
$$
consists of those elements of $W$ which act as cyclic permutations. In particular, it does not contain any reflection.
\qeb

\ebs

Since reflection subgroups can be conveniently described in terms 
of roots, instead of the stabilizers themselves, we will use their maximal reflection 
subgroups to construct a disjoint decomposition of $T^\CC$ into $W$-invariant subsets. We will characterise these subsets in terms of roots and derive defining relations. 
\medskip

For a subgroup $\tilde W$ of $W$, the maximal reflection subgroup of $\tilde W$
is generated by the reflections associated with the elements of the subset 
\beq\label{G-RSS-ReflUGr}
\RSS_{\tilde W} := \{\alpha \in \RS : \sigma_\alpha \in \tilde W\}
\eeq
of $\RS$. Let us describe the subsets of $\RS$ arising this way. Following
\cite{DyerLehrer}, by a root subsystem we mean a subset $\RSS \subset \RS$ which
is a root system in some linear subspace of $\text{span}_\RR\RS$. For
convenience, we include the case of the empty set. It is easy to see that $\RSS$
is a root subsystem iff it is invariant under the reflections associated with
its elements, that is, iff $\sigma_\alpha(\beta) \in \RSS$ for all
$\alpha,\beta\in\RSS$. Thus, as a consequence of Formula \eqref{G-Weyl-Kjg},
$\RSS_{\tilde W}$ is a root subsystem for every subgroup $\tilde W$ of $W$. In
fact, the assignment of \,$\RSS_{\tilde W}$ to $\tilde W$ induces a bijection
between reflection subgroups of $W$ and root subsystems of $\RS$. 

Now, let $x\in T^\CC$ and let $W_x$ denote the stabilizer of $x$ under the
action of $W$. To $x$ we assign the root subsystem 
\beq\label{G-RSSx}
\RSS_x := \RSS_{W_x} \equiv \{\alpha \in \RS : \sigma_\alpha(x) = x\}\,.
\eeq
Due to \eqref{G-Weyl-Kjg}, 
\beq\label{G-RSSwx}
\RSS_{w(x)} = w (\RSS_x) ~~~~~~ \forall ~ w\in W\,.
\eeq
Hence, to the $W$-orbit of $x$ we can associate the conjugacy class of $\RSS_x$ under the action of $W$ on $\RS$. This conjugacy class will be referred to as the reflection type of $x$. It will be denoted by $\rt_x$. 

Now, let $\RT$ denote the set of conjugacy classes of root subsystems and let
$\RTT \subseteq \RT$ denote the subset of reflection types. For $\rt\in\RT$, 
let $T^\CC_\rt$ denote the subset of $T^\CC$ of points having reflection type
$\rt$ and let $\pha_\rt$ to be the subset of $\pha$ of orbits having reflection
type $\rt$. Obviously, $\pha_\rt$ is the image of $T^\CC_\rt$ under the natural
projection $\pr : T^\CC \to \pha$. The subsets $T^\CC_\rt$ and $\pha_\rt$ are 
nonempty precisely iff $\rt \in \RTT$. In this case, we will refer to
them as the reflection type subsets of $T^\CC$ and $\pha$, respectively. As a
result, we obtain disjoint decompositions 
$$
T^\CC = \bigcup_{\rt\in\RTT} T^\CC_\rt
 \,,~~~~~~
\pha = \bigcup_{\rt\in\RTT} \pha_\rt\,.
$$
The set $\RT$ is partially ordered by inclusion modulo conjugacy, i.e., $\rt_1 \leq \rt_2$ if there exist representatives $\RSS_i$ of $\rt_i$ such that $\RSS_1 \subseteq \RSS_2$. Let
$$
T^\CC_\rtt := \{x\in T^\CC : \rt_x \geq \rt\}
 \,,~~~~~~
\pha_\rtt := \pr\left(T^\CC_\rtt\right)\,.
$$
$T^\CC_\rtt$ consists of the points whose stabilizer contains a subgroup which
is generated by a representative of $\rt$. We have the disjoint
decompositions
$$
T^\CC_\rtt = \bigcup_{\rt'\geq\rt} T^\CC_{\rt'}
 \,,~~~~~~
\pha_\rtt = \bigcup_{\rt'\geq\rt} \pha_{\rt'}\,.
$$


\section{Relations for reflection types}
\label{S-Rel}


In this section, we derive relations characterizing the
subset $T^\CC_\rtt$ inside $T^\CC$ for all $\rt\in\RT$.  

In a first step, we determine the root subsystems $\RSS_x$ associated with the
elements of $T^\CC$. Let $x\in T^\CC$. Let $a\in G$ and $B\in\mf t$ such that $x
= a\exp(\mr i B)$ and choose $A\in \mf 
t$ such that $a = \exp(A)$. Then, for $\alpha\in\RS$, we have $\sigma_\alpha(x)
= x$ iff
$$
\exp\big(\sigma_\alpha(A)\big) \exp\big(\mr i \sigma_\alpha(B)\big)
 =
\exp(A) \exp(\mr i B)\,.
$$
By uniqueness of the polar decomposition, this is equivalent to  
$$
\exp\big(\sigma_\alpha(A)\big) = \exp(A)
 ~~~\text{ and }~~~
\sigma_\alpha(B) = B\,.
$$
By \eqref{G-Refl-1}, the second equation holds iff 
$$
\alpha(B) = 0\,.
$$
The first equation holds iff
$$
\sigma_\alpha(A) - A \in\ker(\exp)\,.
$$
To analyse this condition, choose a system of generators $\{H_i\}$ of the
lattice $\ker(\exp)$ and expand 
\beq\label{G-sAA}
\sigma_\alpha(A) - A = \sum_i k_i H_i
\eeq
with integers $k_i$. Since every root is contained in a base, we can find a base $\SSR$ containing $\alpha$. Since $\{\frac{4\pi \mr i}{\kf(\beta,\beta)} H_\beta : \beta \in\SSR\}$ is a basis in $\mf t$ and since the $\ZZ$-lattice
spanned by this basis is contained in the lattice $\ker(\exp)\subset \mf t$, 
there exist rationals $r_{i\beta}$, $i=1,\dots,\rank(G)$, $\beta\in\SSR$, such
that  
$$
H_i = \sum_\beta r_{i\beta} \frac{4\pi \mr i}{\kf(\beta,\beta)} H_\beta
 \,,~~~~~~
i=1,\dots,\rank(G)\,.
$$
Plugging this and \eqref{G-Refl-1} into \eqref{G-sAA}, we obtain the system of equations
\begin{eqnarray}\label{G-GS-qa-1}
2\pi \mr i \sum_i r_{i\alpha} k_i & = & - \alpha(A)\,,
\\ \label{G-GS-qa-2}
2\pi \mr i \sum_i r_{i\beta} k_i & = & 0 ~~~~ \forall ~ \beta \in \SSR\setminus\{\alpha\}\,.
\end{eqnarray}

\ble

There exist relatively prime integers $p_\alpha$, $q_\alpha$ such that the subset
\beq\label{G-GS-qa-Lsg}
\left\{\sum\nolimits_i r_{i\alpha} k_i : (k_1,\dots,k_r) \text{ is an integer solution of \eqref{G-GS-qa-2}}\right\}
\eeq
of $\RR$ coincides with $\frac{p_\alpha}{q_\alpha} \ZZ$.

\ele

\bbw

Denote the subset \eqref{G-GS-qa-Lsg} by $M_\alpha$. Since the integer solutions
of \eqref{G-GS-qa-2} form a group, $M_\alpha$ is a subgroup of $\RR$. Since
there exists an integer $m_\alpha$ such that $m_\alpha r_{\beta i}$ is an
integer for all $i$, $M_\alpha$ is discrete and hence a lattice in $\RR$. Now,
the existence theorem for lattice bases, see e.g.\ \cite[Thm. IX.1.1]{Simon},
yields $y_\alpha\in\RR$ such that $M_\alpha = y_\alpha\ZZ$. Since $M_\alpha$ is
contained in the rationals, we finally obtain $y_\alpha =
\frac{p_\alpha}{q_\alpha}$ for relatively prime integers $p_\alpha$, $q_\alpha$.
\ebw

Thus, we have found that $\sigma_\alpha(x) = x$ iff $\alpha(A) \in 2 \pi \mr i  \frac{p_\alpha}{q_\alpha} \ZZ$ and $\alpha(B) = 0$. Since $\alpha(B)$ is purely imaginary, this condition can be rewritten in the form 
\beq\label{G-sa-Bed}
\mr e^{\frac{q_\alpha}{p_\alpha}\alpha(A+\mr i B)} = 1\,.
\eeq
Recall that every functional $\eta\in(\mf t^\CC)^\ast$ with the property
$\eta\big(\ker(\exp)\big) \subset 2\pi\mr i\ZZ$ descends to a smooth function $e_\eta : T^\CC \to \CC$ via
\beq\label{G-D-e}
e_\eta\big(\exp(X)\big) = \mr e^{\eta(X)}\,.
\eeq
The functional $\frac{q_\alpha}{p_\alpha}\alpha$ is of this type: since
$\ker(\exp) \subset \mf t$ and since it is invariant under $\sigma_\alpha$,
every $X\in\ker(\exp)$ satisfies $\sigma_\alpha(X) - X \in\ker(\exp)$ and hence
$\frac{q_\alpha}{p_\alpha}\alpha(X) \in 2\pi \mr i \ZZ$. Therefore, the function
$e_{\frac{q_\alpha}{p_\alpha}\alpha}$ exists. Define 
$$
\efn\alpha := e_{\frac{q_\alpha}{p_\alpha}\alpha} - 1\,.
$$
Using \eqref{G-sa-Bed} and the fact that $x = \exp(A+\mr i B)$, we finally 
conclude that $\sigma_\alpha(x) = x$ iff $\efn\alpha(x) = 0$. This proves

\bsz\label{S-Gx}

For $x\in T^\CC$, one has $\RSS_x = \{\alpha\in\RS : \efn\alpha(x) = 0\}$.
 \qed

\esz

\bbm\label{Bem-Gx-efzh}

If $G$ is simply connected, $\ker(\exp)$ coincides with the lattice generated by
the elements $2\pi \mr i \, \frac{2 H_\beta}{\kf(\beta,\beta)}$, where $\beta$
belongs to some base in $\RS$. As a consequence, the lattice \eqref{G-GS-qa-Lsg}
coincides with $\ZZ$ and $p_\alpha = q_\alpha = 1$. Hence, in this case,
$\efn\alpha = e_\alpha - 1$. 
\qeb

\ebm

In a second step, we derive relations characterizing the subsets $T^\CC_\rtt$ in
terms of the functions $\efn\alpha$.
For $\rt\in\RT$, choose a representative $\RSS$ and define a function $\rel\rt$
on $T^\CC$ by  
\beq\label{G-D-relrt}
\rel\rt
 := 
 \sum_{w\in W} \prod_{\alpha\in\RS \setminus w(\RSS)} \efn\alpha\,.
\eeq
This function does obviously not depend on the choice of $\RSS$.

\ble\label{L-nr}

For all $x\in T^\CC$, we have $\rel{\rt_x}(x) \neq 0$.

\ele

\bbw

We can choose $\RSS = \RSS_x$ in the definition of $\rel{\rt_x}$. Then, by Proposition \rref{S-Gx}, $\efn\alpha(x) = 0$ for all $\alpha
\in\RSS$ and $\efn\alpha(x) \neq 0$ for all $\alpha\notin\RSS$. Hence, in the sum
over $w\in W$, every term with $w(\RSS) = \RSS$ gives a contribution
$$
\prod_{\alpha\in\RS \setminus w(\RSS)} \efn\alpha(x)
 = 
\prod_{\alpha\in\RS \setminus \RSS} \efn\alpha(x)
$$
which is nonzero and independent of $w$, whereas the terms with $w(\RSS) \neq \RSS$ do not
contribute, because $\RS \setminus w(\RSS)$ contains some $\alpha\in\RSS$ here.
Consequently, $\rel{\rt_x}(x)$ is a multiple of a nonzero complex number. 
\ebw

\btm\label{T-T(r)}\label{T-RT}

Let ${\rt_0}\in\RT$ and $x\in T^\CC$. Then, $x\in T^\CC_{(\rt_0)}$ iff $\rel{\rt}(x)
= 0$ for all $\rt \not\geq {\rt_0}$.

\etm

As a consequence, 
\beq\label{G-TC(r)}
T^\CC_{(\rt_0)} = \bigcap_{\rt \not\geq {\rt_0}} \rel{\rt}^{-1}(0)\,.
\eeq

\bbw

First, assume that $x\in T^\CC_{(\rt_0)}$. Let $\rt\in\RT$ such that
$\rt\not\geq{\rt_0}$ and choose a representative $\RSS$ of $\rt$ to compute
$\rel{\rt}$ according to \eqref{G-D-relrt}. Since $\rt_x \geq \rt_0$, we have
$\rt\not\geq\rt_x$. This means that for any $w\in W$, $\RSS_x$ is not a subset
of $w(\RSS)$ and hence $\RS \setminus w(\RSS)$ contains an element of $\RSS_x$.
Thus, Proposition \rref{S-Gx} implies that $\rel{\rt}(x) = 0$. 
Conversely, assume that $\rel{\rt}(x) = 0$ for all $\rt\in\RT$ such that $\rt
\not\geq {\rt_0}$. Then, Lemma \rref{L-nr} implies $\rt_x \geq {\rt_0}$ and
hence $x\in T^\CC_{(\rt_0)}$. 
\ebw

Theorem \rref{T-T(r)} yields the following characterization of $T^\CC_\rt$.

\bfg\label{F-Tr}

Let ${\rt_0}\in\RT$ and $x\in T^\CC$. Then, $x\in T^\CC_{\rt_0}$ iff $\rel{\rt_0}(x)\neq0$ and $\rel{\rt}(x) = 0$ for all $\rt \not\geq {\rt_0}$.

\efg

\bbw

If $x\in T^\CC_{\rt_0}$, then Theorem \rref{T-T(r)} implies that $\rel{\rt}(x) =
0$ for all $\rt\not\geq{\rt_0}$ and Lemma \rref{L-nr} implies that
$\rel{\rt_0}(x) = \rel{\rt_x}(x) \neq 0$.  
Conversely, if $\rel{\rt}(x) = 0$ for all $\rt\not\geq{\rt_0}$, the theorem
yields $x\in T^\CC_{(\rt_0)}$. If, in addition, $\rel{\rt_0}(x)\neq 0$, then $x$
cannot belong to $T^\CC_{(\rt')}$ for any $\rt' > {\rt_0}$, because $\rt' >
{\rt_0}$ implies ${\rt_0}\not\geq\rt'$, so that every $y\in T^\CC_{(\rt')}$
satisfies $\rel{\rt_0}(y)=0$. Thus, $x\in T^\CC_{(\rt_0)} \setminus
\left(\bigcup_{\rt' > {\rt_0}} T^\CC_{(\rt')}\right) = T^\CC_{\rt_0}$. 
\ebw

\bbm\label{Bem-T(r)}

Let $\rt,\rt'\in\RT$ and assume that $\rt'$ is a direct successor of $\rt$.
According to Theorem \rref{T-T(r)}, the subset $T^\CC_{(\rt')}$ of
$T^\CC_{(\rt)}$ is the set of solutions of the relations $\rel\rt=0$ and
$\rel{\rt''}=0$ for all $\rt'' > \rt$ such that $\rt''\not\geq\rt'$ (that is,
all successors of $\rt$ which cannot be compared with $\rt'$). Of course, there
might be further reflection types to be taken into account to define
$T^\CC_{(\rt')}$ as a subset of $T^\CC$. From the slice theorem we know that
strata whose orbit types are not comparable are separated, i.e., they admit open
neighbourhoods which do not intersect. This carries over to reflection type
subsets, because the decomposition by orbit types is at least as fine as that by
reflection types. Thus, the relation $\rel\rt=0$ defines $T^\CC_{(\rt')}$
locally, i.e., as a subset of some open subset of $T^\CC_{(\rt)}$, whereas the
relations $\nu_{\rt''}=0$ for all $\rt''>\rt$ such that $\rt''\not\geq\rt'$ are
necessary to define $T^\CC_{(\rt')}$ globally as a subset of $T^\CC_{(\rt)}$. In
particular, they do not reduce the dimension. Note that the relation $\rel\rt=0$
reduces the dimension iff $\rt$ is infinitesimal, cf.\ Remark \rref{Bem-RT-HO}
below. 
\qeb

\ebm

In a third step, we express the functions $\rel\rt$ in terms of the normalized
characters $\ncc_\hw$ of $G^\CC$, see Section \rref{A-adjQuo}.
For that purpose, we choose a Weyl chamber in $(\mf t^\CC)^\ast$ and denote the
corresponding sets of positive roots and of dominant weights by $\RS_+$ and
$\HW$, respectively. Recall that for every functional $\eta\in(\mf t^\CC)^\ast$
with the property $\eta\big(\ker(\exp)\big) \subset 2\pi\mr i\ZZ$, the function
$e_\eta$ is defined by \eqref{G-D-e}. Let 
\beq\label{G-D-Ad}
A_\eta := \sum_{w\in W} \sign(w) \, e_{w\eta}\,.
\eeq
Let $\rt\in\RT$. We expand
$$
\rel\rt = \sum_{\hw \in \HW} C^\rt_\hw \, \ncc_\hw\,.
$$
Since restriction to $G$ yields 
$$
(\rel\rt)_{\upharpoonright G}
 = 
\sum_{\hw \in \HW} \frac{C^\rt_\hw}{N_\hw} \, \chi_\hw\,,
$$
we can compute the coefficients $C^\rt_\hw$ via the integrals
$$
C^\rt_\hw = N_\hw \int_G \ol{\chi_\hw} \,\, \rel\rt \, 	\mr d\mu\,.
$$
By Weyl's integration formula,
$$
C^\rt_\hw = \frac{N_\hw}{|W|} \int_T \ol{\chi_\hw} \,\, \rel\rt \, 
|A_\delta|^2 \mr d\mu\,,
$$
where $\delta$ denotes half the sum of the positive roots, given by
\eqref{G-D-delta}, $|W|$ denotes the cardinality (order) of $W$ and $\mr d\mu_T$ denotes the normalized Haar measure on $T$. Using Weyl's character formula,
$$
\chi_\hw = \frac{A_{\hw + \delta}}{A_\delta}\,,
$$
we can rewrite this as 
\beq\label{G-CharZlg-1}
C^\rt_\hw
 = 
\frac{N_\hw}{|W|} \int_T \ol{A_{\hw + \delta}} \, A_\delta \, \rel\rt  \, \mr d\mu_T\,.
\eeq
To evaluate the integral, we choose a root subsystem $\RSS$ representing $\rt$ and expand
\begin{align*}
\rel\rt
 =
\sum_{w\in W} \prod_{\alpha\in\RS\setminus w(\RSS)} \efn{\frac{q_\alpha}{p_\alpha}\alpha}
 =
\sum_{w\in W} \prod_{\alpha\in\RS\setminus\RSS} 
\left(e_{\frac{q_\alpha}{p_\alpha}w\alpha} - 1\right)
 =
\sum_{w\in W} \sum_{\CSS\subseteq\RS\setminus\RSS} (-1)^{|\RS\setminus\RSS| - |\CSS|} \prod_{\alpha\in\Delta} e_{\frac{q_\alpha}{p_\alpha} w \alpha}\,,
\end{align*}
where $|\cdot|$ denotes the cardinality. Since both $\RS$ and $\RSS$ are root
systems, the 
cardinality of $\RS\setminus\RSS$ is even. Thus, using $e_{\eta_1} e_{\eta_2} =
e_{\eta_1 + \eta_2}$, as well as the notation 
$$
\hw_\CSS := \sum_{\alpha\in\CSS} \frac{q_\alpha}{p_\alpha} \alpha\,,
$$
with $\hw_\emptyset = 0$ understood, we obtain
\beq\label{G-relrt-e}
\rel\rt
 = 
\sum_{w\in W} \sum_{\CSS\subseteq\RS\setminus\RSS} (-1)^{|\CSS|} \, e_{w\hw_\CSS}\,.
\eeq
Plugging this and \eqref{G-D-Ad} into \eqref{G-CharZlg-1} and using that on $G$ one has $\ol{e_\eta} = e_{-\eta}$, we arrive at
$$
C^\rt_\hw
 = 
\frac{N_\hw}{|W|} 
 \sum\nolimits_{w,w',w''} \sum\nolimits_{\CSS\subseteq\RS\setminus\RSS} 
(-1)^{|\CSS|} \, \sign(w') \, \sign(w'')
 \int_T 
e_{w\hw_\CSS + w' \delta - w''(\hw + \delta)} \,\, \mr d\mu_T
$$
The integral vanishes unless $w\hw_\CSS + w' \delta - w''(\hw + \delta) = 0$, in which case it evaluates to $1$. Hence, we can write
$$
C^\rt_\hw
 = 
\frac{N_\hw}{|W|} 
 \sum\nolimits_{w,w',w''} \sum\nolimits_{\CSS\subseteq\RS\setminus\RSS} 
(-1)^{|\CSS|} \, \sign(w') \, \sign(w'') \,\,
\updelta_{w\hw_\CSS , w''(\hw + \delta) - w' \delta}\,,
$$
where upright $\updelta$ denotes the Kronecker symbol. By shifting the summation over $w'$ by $w^{-1}$ and the summation over $w''$ by $w^{-1}w'^{-1}$, and by carrying out the sum over $w$, we obtain 
$$
C^\rt_\hw
 = 
N_\hw \sum\nolimits_{w',w''} \sum\nolimits_{\CSS\subseteq\RS\setminus\RSS} 
(-1)^{|\CSS|} \, \sign(w'') \,\,
\updelta_{\hw_\CSS , w'(w''(\hw + \delta) - \delta)}\,.
$$
In effect, the sum over $\CSS$ counts the number of subsets $\CSS \subseteq \RS\setminus\RSS$ which satisfy $\hw_\CSS = w'(w''(\hw + \delta) - \delta)$, weighted by a factor $(-1)^{|\CSS|}$. Thus, for $\eta\in(\mf t^\CC)^\ast$, let $\vfhp\RSS(\eta)$ denote the number of subsets $\CSS \subseteq \RS\setminus\RSS$ which satisfy $\hw_\CSS = \eta$ and which have an even number of elements (including the empty set) and let $\vfhm\RSS(\eta)$ denote the corresponding number of subsets which have an odd number of elements. Define
$$
\vfh\RSS(\eta) = \vfhp\RSS(\eta) - \vfhm\RSS(\eta)\,.
$$
Then,
\beq\label{G-Cl}
C^\rt_\hw
 = 
N_\hw \sum_{w',w''\in W} \sign(w'') \, \vfh\RSS\big(w'(w''(\hw + \delta) - \delta)\big)\,.
\eeq
Let $\HW^\rt \subset \HW$ denote the subset consisting of elements $\hw$ such that $C^\rt_\hw \neq 0$.

\bsz\label{S-Rel-Char}

For $\rt\in\RT$,
$$
\rel\rt
 = 
\sum_{\hw\in\HW^\rt} C^\rt_\hw \, \ncc_\hw
 \,,~~~~~~
C^\rt_\hw
 = 
N_\hw \sum_{w',w''\in W} \sign(w'') \, \vfh\RSS\big(w'(w''(\hw + \delta) - \delta)\big)\,,
$$
where $\RSS$ is a root subsystem representing $\rt$.
 \qed

\esz

The coefficients $C^\rt_\hw$ represent $\rel\rt$ under the identification of $\mc H$ with the Hilbert space $\ell^2$ over the index set $\HW$, see Section \rref{A-adjQuo}. They can be easily computed using computer algebra, see Section \rref{A-ex}. This concludes the discussion of the relations characterizing the subsets $T^\CC_\rtt \subseteq T^\CC$.

\bbm\label{Bem-Cl}

Depending on $\RSS$, the computation of the coefficients $C^\rt_\hw$ can be
simplified by rewriting the sum over $w'$ in \eqref{G-Cl} as
follows. Let  
$$
W_\RSS := \{ w\in W : w(\RSS) = \RSS\}\,.
$$
If $w'\big(w''(\hw + \delta) - \delta\big) = \lambda_\CSS$ and $w\in W_\RSS$, then $ww'\big(w''(\hw + \delta) - \delta\big) = \lambda_{w\CSS}$, where $w\CSS \subset \RS\setminus \RSS$, and $\vfh\RSS\big(ww'(w''(\hw + \delta) - \delta)\big) = \vfh\RSS\big(w'(w''(\hw + \delta) - \delta)\big)$. Hence, using the notation 
$$
\tvfh\RSS(\eta)
 :=
\sum_{w'\in W/W_\RSS}\, \vfh\RSS\big(w'(\eta)\big)\,,
$$
where the sum runs over a selection of one representative for each right $W_\RSS$-coset,
Formula \eqref{G-Cl} can be rewritten as $C^\rt_\hw = |W_\RSS| \tilde C^\rt_\hw$ with the reduced coefficients
\beq\label{G-Cl-red}
\tilde C^\rt_\hw
 = 
N_\hw \sum_{w''\in W} \sign(w'') \, \tvfh\RSS\big(w''(\hw + \delta) - \delta\big)
\eeq
and instead of the functions $\rel\rt$ one may work with the reduced functions $\rel\rt/|W_\RSS|$. The multiplicities $\tvfh\RSS(\eta)$ count the number of ways of writing $\eta$ in the form $\eta = w'^{-1}(\hw_\CSS)$ with $w'\in W/W_\RSS$ and with $\CSS$ being a subset of $\RS \setminus \RSS$ with an even number of elements minus this number with $\CSS$ having an odd number of elements. 
\qeb

\ebm

\bbs\label{B-Cl-SU2}

Let us illustrate the formula for $C^\rt_\hw$ given in Proposition \rref{S-Rel-Char} for $G=\SU(2)$. The root system is $\RS = \{\alpha,-\alpha\}$, the Weyl group is $W = \{1,-1\}$, acting by multiplication, and $\delta = \frac\alpha2$. Dominant weights are of the form $\hw_k = \frac k 2 \alpha$ with a nonnegative integer $k$ and 
$$
w'\big(w''(\hw_k+\delta) - \delta\big)
 =
\begin{cases}
\pm \frac k 2 \alpha &|~ w' = \pm 1\,,~ w'' = 1\,,
\\
\pm \frac{k+2} 2 \alpha &|~ w' = \pm 1\,,~ w'' = - 1\,.
\end{cases}
$$
Since $\SU(2)$ is simply connected, $q_\alpha = p_\alpha = 1$.
There are two root subsystems, $\RSS = \emptyset$ and $\RSS = \RS$. We have $T^\CC_{([\emptyset])} = T^\CC$ and $T^\CC_{([\RS])} = T^\CC_{[\RS]}$. Since $[\emptyset] \not \geq [\RS]$, the subset $T^\CC_{[\RS]} \subset T^\CC$ is defined by the relation $\rel{[\emptyset]} = 0$. To determine $\rel{[\emptyset]}$, we calculate $\lambda_\Delta$ for all subsets $\Delta \subseteq \RS \setminus \emptyset = \RS$:
$$
\lambda_\Delta
 = 
\begin{cases}
\alpha & \Delta = \{\alpha\}\,,
\\
-\alpha & \Delta = \{-\alpha\}\,,
\\
0 & \Delta = \emptyset \text{ or } \{\alpha,-\alpha\}\,.
\end{cases}
$$
Hence,
$$
V_{\emptyset}(\pm\alpha) = -1
\,,~~~~~~
V_{\emptyset}(0) = 2
$$
and $V_{\emptyset}(\eta) = 0$ for all other $\eta$. For given $k$, we compute
$$
w'\big(w''(\hw_k+\delta) - \delta\big)
 =
w' \frac{w''(k+1) - 1}{2} \alpha
 =
\begin{cases}
\pm \frac k 2 \alpha &|~ w' = \pm 1\,,~ w'' = 1\,,
\\
\pm \frac{k+2} 2 \alpha &|~ w' = \pm 1\,,~ w'' = - 1\,.
\end{cases}
$$
Hence, taking the sum over $w',w''$ in the order $(w',w'') = (1,1)$, $(-1,1)$, $(1,-1)$, $(-1,-1)$, we obtain
$$
\frac{C^{[\emptyset]}_{\lambda_0}}{N_{\lambda_0}}
 =
V_\RSS(0) + V_\RSS(0) - V_\RSS(-\alpha) - V_\RSS(\alpha)
 =
6
 \,,~~~~~~
\frac{C^{[\emptyset]}_{\lambda_2}}{N_{\lambda_2}}
 =
V_\RSS(\alpha) + V_\RSS(-\alpha)
 =
-2\,.
$$
All other coefficients vanish. Since $\RSS = \emptyset$ is invariant under all
of $W$, according to Remark \rref{Bem-Cl}, the corresponding reduced
coefficients are 
$$
\frac{\tilde C^0_{\lambda_0}}{N_{\lambda_0}} = 3
 \,,~~~~~~
\frac{\tilde C^0_{\lambda_2}}{N_{\lambda_2}} = -1\,.
$$

Let us add that under the adjoint action of $\SU(2)$, the stabilizers of the
elements of $T^\CC_{[\emptyset]}$ coincide with the center and the stabilizers
of the elements of $T^\CC_{[\RS]}$ coincide with $\SU(2)$. Hence, in the
present case, reflection types and orbit types are equivalent. This is true for
all $\SU(n)$, see Remark \rref{Bem-SU(n)}. 
\qeb

\ebs


\section{Costratification}
\label{S-kostrfHiR}


In this section, we sketch how to use the relations derived above to construct
the subspaces $\mc H_\rt$ associated with the reflection types. 

Consider a chosen reflection type ${\rt_0}$. According to Theorem \rref{T-RT}, then $T_{(\rt_0)}^\CC$ is defined by the relations $\rel{\rt}=0$ for all $\rt\not\geq{\rt_0}$. For every $\rt\in\RT$, multiplication by $\rel{\rt}$ defines a linear operator
$$
K^{\rt} : \mc H \to \mc H \,,~~~~~~ K^{\rt} \psi = \rel{\rt}\psi\,,
$$
which can be easily seen to be bounded and to have closed range. Then, 
$$
\mc V_{\rt_0} = \sum_{\rt\not\geq{\rt_0}} K^{\rt}(\mc H)
$$
and, consequently,
\beq\label{G-Hrt}
\mc H_{\rt_0} = \bigcap_{\rt\not\geq{\rt_0}} \ker\big((K^{\rt})^\dagger\big)\,.
\eeq
Consider the matrix elements 
$$
K^{\rt}_{\lambda'\lambda} = \langle \ncc_{\lambda'} , K^{\rt} \ncc_\lambda \rangle
$$
of $K^{\rt}$ with respect to the orthonormal basis of normed complex characters
$\ncc_\lambda$. Since $\rel{\rt}$, as well as any product of normed complex
characters, is a {\em real} linear combination of normed complex characters, the
matrix elements $K^{\rt}_{\lambda'\lambda}$ are real. In terms of these matrix
elements, the condition that a function $\psi = \sum_{\lambda\in\HW} H_\lambda
\ncc_\lambda$ belongs to $\mc H_{\rt_0}$ reads 
\beq\label{G-Bed-Hrt}
\sum_{\lambda'\in\HW} K^{\rt}_{\lambda'\lambda} H_{\lambda'} = 0
 ~~~~~~
\forall ~ \lambda\in\HW \,,~ \rt \not\geq \rt_0\,.
\eeq
To compute the matrix elements $K^{\rt}_{\lambda'\lambda}$, we use Proposition \rref{S-Rel-Char} and the expansion 
\beq\label{G-ncc-Pr}
\ncc_{\lambda'} \ncc_\lambda
 = 
\sum_{\lambda''\in\HW} M_{\lambda'\lambda\lambda''} \ncc_{\lambda''}
\eeq
to obtain
\beq\label{G-Kll-CD}
K^{\rt}_{\lambda'\lambda}
 =
\sum_{\lambda''\in\HW^{\rt}} C^{\rt}_{\lambda''} M_{\lambda''\lambda\lambda'}\,.
\eeq
The coefficients $C^{\rt}_{\lambda''}$ are given by Proposition \rref{S-Rel-Char}. For the coefficients $M_{\lambda''\lambda\lambda'}$, the representation theory of semisimple Lie algebras yields the following. By restricting the expansion \eqref{G-ncc-Pr} to $G$ and renaming the labels $\lambda',\lambda,\lambda''$, we obtain the equation
$$
\chi_{\lambda''} \chi_{\lambda}
 = 
\sum_{\lambda'\in\HW} 
\frac{N_{\lambda''} N_\lambda}{N_{\lambda'}} M_{\lambda''\lambda\lambda'} 
 \, 
\chi_{\lambda'}\,.
$$
For $\lambda''\in\HW$, let $\WS(\lambda'')$ denote the weight system of the corresponding irrep. For a weight $\mu\in\WS(\lambda'')$ of this irrep, let $m_{\lambda''}(\mu)$ denote the multiplicity. Finally, for $\lambda,\lambda'\in\HW$ and a weight $\mu$ of an arbitrary irrep, there exists at most one $w \in W$ such that 
\beq\label{G-D-tll}
w(\lambda + \mu + \delta) = \lambda' + \delta\,.
\eeq
Let $\tau_{\lambda\lambda'}(\mu) = \sign(w)$ in case such $w$ exists and $\tau_{\lambda\lambda'}(\mu) = 0$ otherwise. Then,
$$
M_{\lambda''\lambda\lambda'}
 =
\frac{N_{\lambda'}}{N_\lambda N_{\lambda''}}
\sum_{\mu\in\WS(\lambda'')} m_{\lambda''}(\mu) \, \tau_{\lambda\lambda'}(\mu)\,.
$$
Since the multiplicities $m_{\lambda''}(\mu)$ are invariant under $W$ and since every $W$-orbit in $\WS(\lambda'')$ intersects $\HW$ in exactly one point \cite{MoodyPatera}, the sum over $\mu\in\WS(\lambda'')$ may be split as follows. Let 
$$
\HWS(\lambda'') := \WS(\lambda'') \cap \HW
$$
be the subset of dominant weights in the irrep $\lambda''$. For a dominant weight $\mu\in\HWS(\lambda'')$, define
$$
T_{\lambda\lambda'}(\mu)
 := 
\sum_{\mu'\in W(\mu)} \tau_{\lambda\lambda'}(\mu')\,,
$$
where $W(\mu)$ denotes the $W$-orbit of $\mu$. Then,
$$
M_{\lambda''\lambda\lambda'}
 =
\frac{N_{\lambda'}}{N_{\lambda}N_{\lambda''}}
\sum_{\mu\in\HWS(\lambda'')} m_{\lambda''}(\mu) \, T_{\lambda\lambda'}(\mu)\,.
$$
Plugging this into \eqref{G-Kll-CD}, we obtain
$$
K^{\rt}_{\lambda'\lambda}
 =
\frac{N_{\lambda'}}{N_\lambda}
 \sum_{\lambda''\in\HW^{\rt}} 
\frac{C^{\rt}_{\lambda''}}{N_{\lambda''}}
 \sum_{\mu\in\HWS(\lambda'')}
m_{\lambda''}(\mu) \, T_{\lambda\lambda'}(\mu)\,.
$$
To separate the irregular factor $T_{\lambda\lambda'}(\mu)$, we interchange the order of summation. Let 
$$
\HWS^{\rt} := \bigcup_{\lambda''\in\HW^{\rt}} \HWS(\lambda'')
$$
(the dominant weights of all the irreps in $\HW^{\rt}$). Then, 
\beq\label{G-Kll}
K^{\rt}_{\lambda'\lambda}
 =
\frac{N_{\lambda'}}{N_\lambda}
 \sum_{\mu\in\HWS^{\rt}}
D_\mu^{\rt} \, T_{\lambda\lambda'}(\mu)
 ~~ \text{ with } ~~
D_\mu^{\rt}
 :=
\sum_{\lambda''\in\HW^{\rt}}
\frac{C^{\rt}_{\lambda''}}{N_{\lambda''}} \, m_{\lambda''}(\mu)\,,
\eeq
with $m_{\lambda''}(\mu) = 0$ if $\mu\notin\WS(\lambda'')$ understood. The factors $D^{\rt}_\mu$ can be easily determined from Proposition \rref{S-Rel-Char} and Freudenthal's formula for the multiplicities $m_{\lambda''}(\mu)$, see the examples in Section \rref{A-ex}. The factors $T_{\lambda\lambda'}(\mu)$ are irregular and have to be determined case by case. However, if $\lambda$ is sufficiently far from the boundary of the chosen dominant Weyl chamber, these factors become very simple. Let us call $\lambda\in\HW$ stable if $\lambda + \mu$ belongs to the dominant Weyl chamber for all $\mu\in\HWS^{\rt}$. For stable $\lambda$ and arbitrary $\lambda'$, there exists $w\in W$ such that \eqref{G-D-tll} holds iff $\lambda + \mu' = \lambda'$. Thus, $\tau_{\lambda\lambda'}(\mu') = \updelta_{\lambda+\mu',\lambda'}$ and hence
$$
T_{\lambda\lambda'}(\mu)
 = 
 \begin{cases} 
1 &|~ \lambda' - \lambda \in W(\mu)\,,
\\
0 &|~ \text{otherwise.}
 \end{cases}
$$
Plugging this into \eqref{G-Kll} and using that the $W$-orbits of distinct elements of $\HWS^{\rt}$ do not intersect, we obtain the simple formula
\beq\label{G-Kll-stable}
K^{\rt}_{\lambda+w(\mu),\lambda}
 =
 \begin{cases} 
\frac{N_{\lambda+w(\mu)}}{N_\lambda} \, D_\mu^{\rt} & | ~ \mu\in\HWS^{\rt} , w\in W,
\\
0 &|~ \text{otherwise,}
 \end{cases}
\eeq
which holds for stable $\lambda\in\HW$.

\bbm\label{Bem-Kll}

All of the above can be worked out equally well by means of the reduced functions $\rel\rt/|W_\RSS|$, discussed in Remark \rref{Bem-Cl}. This leads to a reduced operator $\tilde K^\rt = K^\rt / |W_\RSS|$ and to reduced coefficients $\tilde D^\rt_\mu = D^\rt_\mu / |W_\RSS|$. Formulae \eqref{G-Kll} and \eqref{G-Kll-stable} remain true if one replaces $K^\rt$ by $\tilde K^\rt$, $C^\rt_\hw$ by $\tilde C^\rt_\hw$ and $K^\rt$ by $\tilde K^\rt_\hw$.
\qeb

\ebm

\bbs\label{B-Kll-SU2}

To illustrate Formula \eqref{G-Kll} for the coefficients $D^\rt_\mu$, consider $G = \SU(2)$. Recall from Example \rref{B-Cl-SU2} that the only root subsystem different from $\RS$ is $\RSS = \emptyset$, defining the generic reflection type $\rt=0$. We use the notation and the results of that example. The elements of $\HW^0$ are $\hw'' = \hw_0,\hw_2$. The dominant weights of the corresponding irreps are $\hw_0$ for $\hw'' = \hw_0$ and $0,\hw_2$ for $\hw'' = \hw_2$. The multiplicities of these weights are $1$. Hence, \eqref{G-Kll} yields 
$$
D^0_{\lambda_0}
 = 
\frac{C^0_{\lambda_0}}{N_{\lambda_0}} m_{\lambda_0}(\lambda_0)
 + 
\frac{C^0_{\lambda_2}}{N_{\lambda_2}} m_{\lambda_2}(\lambda_0)
 =
4
 \,,~~~~~~
D^0_{\lambda_2}
 = 
\frac{C^0_{\lambda_2}}{N_{\lambda_2}} m_{\lambda_2}(\lambda_2)
 =
-2\,.
$$
Since $\RSS$ is invariant under all of $W$ and $|W| = 2$, the corresponding reduced coefficients are $\tilde D^0_{\lambda_0} = 2$, $\tilde D^0_{\lambda_2} = -1$.
\qeb

\ebs


\section{Root subsystems}
\label{A-RSS}


To complete the discussion of the reflection type decomposition, we still have
to find the set $\RT$ of $W$-conjugacy classes of root subsystems of $\RS$.
The classification of root subsystems up to the action of $W$ or, equivalently,
the classification of reflection subgroups of $W$ up to conjugacy under inner
automorphisms, can be found in \cite{Douglass,DyerLehrer,FeliksonTumarkin}. See
\S 1.3 of \cite{DyerLehrer} for historical remarks. 

Following \cite{DyerLehrer}, we say that a root subsystem $\RSS$ of $\RS$ is
closed if for all $\alpha, \beta \in \RSS$ such that $\alpha + \beta \in \RS$
one has $\alpha + \beta \in \RSS$. Beware that in \cite{FeliksonTumarkin}, the
term root subsystem is used to refer to what is here called a closed root
subsystem. Obviously, the linear subspace 
$$
\mf g^\CC_\RSS
 := 
\text{span}_\CC\{H_\alpha : \alpha \in \RSS\}
 \oplus
\bigoplus_{\alpha\in\RSS} \mf g^\CC_\alpha\,,
$$
of $\mf g^\CC$ is a Lie subalgebra iff $\RSS$ is closed. Because of this connection with Lie subalgebras and Lie subgroups, the closed root subsystems have been classified up to conjugacy under $W$ long ago by Borel and de Siebenthal \cite{Borel} and Dynkin \cite{Dynkin}. Their analysis is based on the following observation.
Recall that a subset $\SSR \subset \RS$ is called a base (or a simple system) of
$\RS$ if it is a basis in $(\mf t^\CC)^\ast$ with respect to which the elements
of $\RS$ have either positive or negative integer coefficients. Given a closed
root subsystem $\RSS$ and a base $\SSR$ in $\RSS$, let $\SSR'$ be obtained by
extending $\SSR$ by the negative of the highest (with respect to $\SSR$) root of
$\RSS$. Then, every proper subset of $\SSR'$ is a base of a closed root
subsystem of $\RSS$, and hence the base of a closed root subsystem of $\RS$.
This way, up to conjugacy under $W$, all the closed root subsystems of $\RS$ can
be generated iteratively from a base in $\RS$. The remaining, non-closed, root
subsystems arise by a similar algorithm
\cite{Carter,Douglass,DyerLehrer,FeliksonTumarkin}: 
as usual, define the dual roots by 
$$
\alpha^\vee := \frac{2\alpha}{k(\alpha,\alpha)}
 \,,~~~~~~
\alpha\in\RS\,.
$$
Given $\RSS$ and $\SSR$, let $\SSR''$ be obtained by extending $\SSR$ by the negative of the dual of the highest (with respect to $\SSR^\vee$) root of the dual root system $\RSS^\vee$. Then, every proper subset of $\SSR''$ is a base of a root subsystem of $\RSS$, and hence the base of a root subsystem of $\RS$. Up to conjugacy under $W$, the root subsystems generated iteratively in this way from a base of $\RS$ include all the non-closed root subsystems of $\RS$. Since $\RS^\vee$ differs from $\RS$ only if it contains roots of different lengths, this implies in particular that non-closed root subsystems can exist only in root systems containing a factor $B_l$, $C_l$, $F_4$ or $G_2$. In contrast, all the root subsystems of $A_l$, $D_l$, $E_6$, $E_7$ and $E_8$ are closed. 

For the classical series, the algorithm described above yields the following result. Given a base $\SSR = \{\alpha_1,\dots,\alpha_n\}$ in $\RS$, denote 
 \ala{
\alpha^A_l & := \alpha_l + \dots + \alpha_n\,,
 &
\alpha^B_l & := \alpha_l + 2( \alpha_{l+1} + \dots + \alpha_n)\,, 
\\
\alpha^C_l & := 2(\alpha_l + \dots + \alpha_{n-1}) + \alpha_n\,,
 &
\tilde\alpha^C_l & := \alpha_l + 2(\alpha_{l+1} + \dots + \alpha_{n-1}) + \alpha_n\,,
 }
where $1 \leq l < n$ and 
$$
\alpha^D_l
 := 
\alpha_l + 2(\alpha_{l+1} + \dots + \alpha_{n-2}) + \alpha_{n-1} + \alpha_n\,,
$$
where $1 \leq l < n-2$. Given a sequence $(i_1,\cdots,i_n)$, let $(i^1,\dots,i^n)$ denote the sequence of partial sums, i.e., $i^1 = i_1$, $i^2 = i_1 + i_2$ etc.

\bsz\label{S-RT-klGr} 

The following list contains the $W$-conjugacy classes of root subsystems of the classical series and a base of a representative for each class. Assume $n \geq 1$ for $A_n$, $n \geq 2$ for $B_n$, $C_n$ and $n\geq 4$ for $D_n$, as well as $i_l,k_l \geq 1$ and $j_l \geq 2$ for all $l$.

\ben
 
\item[$(A_n)$] 

Classes:\abs $A_{i_1} \oplus \cdots \oplus A_{i_r}$, where $r \geq 0$, $i_1 \leq \cdots \leq i_r$ and $i^r + r - 1 \leq n$. 

Bases:\abs\abs
\unitlength0.75cm
\begin{picture}(10,0.75)

 \put(0,0.1){
\Ateil{0,0}
\Ateil{3,0}
\Pure{5.5,0}
\Ateil{7,0}

\marke{0,0.25}{bc}{\alpha_1}
\marke{2,0.25}{bc}{\alpha_{i_1}}
\marke{3,0.25}{bc}{\alpha_{i_1+2}}
\marke{5,0.25}{bc}{\alpha_{i^2+1}}
\marke{7.25,0.25}{bc}{\alpha_{i^{r\minus1}+r}}
\marke{9.75,0.25}{bc}{\alpha_{i^r+r-1}}
 }
\end{picture}

All root subsystems are closed.

\item[$(B_n)$] 

Classes:\abs 
$
A_{i_1} \oplus \cdots \oplus A_{i_r} 
 \oplus 
D_{j_1} \oplus \cdots \oplus D_{j_s} 
 \oplus 
B_{k_1} \oplus \cdots \oplus B_{k_t}
$,
where $r,s,t \geq 0$, $i_1 \leq \cdots \leq i_r$, $j_1 \leq \cdots \leq j_s$, $k_1 \leq \cdots \leq k_{t-1}$ and $i^r + r + j^s + k^t \leq n$.

Bases:

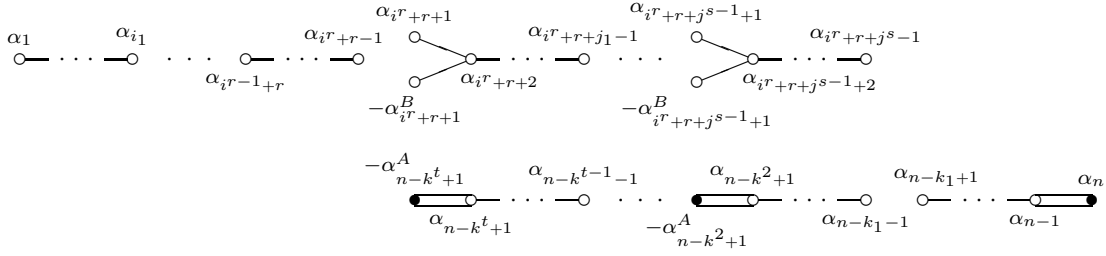
\unitlength0.75cm

\begin{picture}(19,4)

 \put(0,3){

\Ateil{0,0}
\Pure{2.5,0}
\Ateil{4,0}

\Dre{7,0}\Ateil{8,0}
\Pure{10.5,0}
\Dre{12,0}\Ateil{13,0}

\marke{0,0.25}{bc}{\alpha_1}
\marke{2,0.25}{bc}{\alpha_{i_1}}
\marke{4,-0.25}{tc}{\alpha_{i^{r \minus 1} \plus r}}
\marke{5.75,0.25}{bc}{\alpha_{i^r \plus r \minus 1}}

\marke{7,0.65}{bc}{\alpha_{i^r \plus r \plus 1}}
\marke{7,-0.65}{tc}{-\alpha^B_{i^r \plus r \plus 1}}
\marke{8.5,-0.25}{tc}{\alpha_{i^r \plus r \plus 2}}
\marke{10,0.25}{bc}{\alpha_{i^r \plus r \plus j_1 \minus 1}}

\marke{12,0.65}{bc}{\alpha_{i^r \plus r \plus j^{s-1} \plus 1}}
\marke{12,-0.65}{tc}{-\alpha^B_{i^r \plus r \plus j^{s-1} \plus 1}}
\marke{14,-0.25}{tc}{\alpha_{i^r \plus r \plus j^{s-1} \plus 2}}
\marke{15,0.25}{bc}{\alpha_{i^r \plus r \plus j^s \minus 1}}

 }

 \put(7,0.5){

\Bre{0,0}\Ateil{1,0}
\Pure{3.5,0}
\Bre{5,0}\Ateil{6,0}
\Ateil{9,0}
\Bli{12,0}

\marke{0,0.25}{bc}{-\alpha^A_{n \minus k^t \plus 1}}
\marke{1,-0.25}{tc}{\alpha_{n \minus k^t \plus 1}}
\marke{3,0.25}{bc}{\alpha_{n \minus k^{t \minus 1} \minus 1}}

\marke{5,-0.25}{tc}{-\alpha^A_{n \minus k^2 \plus 1}}
\marke{6,0.25}{bc}{\alpha_{n \minus k^2 \plus 1}}
\marke{8,-0.25}{tc}{\alpha_{n \minus k_1 \minus 1}}

\marke{9.25,0.25}{bc}{\alpha_{n \minus k_1 \plus 1}}
\marke{11,-0.25}{tc}{\alpha_{n \minus 1}}
\marke{12,0.25}{bc}{\alpha_n}

 }

\end{picture}

Closed root subsystems are those with $t \leq 1$.

\item[$(C_n)$] Classes:\abs 
$
A_{i_1} \oplus \cdots \oplus A_{i_r} 
 \oplus 
D_{j_1} \oplus \cdots \oplus D_{j_s} 
 \oplus 
C_{k_1} \oplus \cdots \oplus C_{k_t}
$,
where $r,s,t \geq 0$, $i_1 \leq \cdots \leq i_r$, $j_1 \leq \cdots \leq j_s$, $k_1 \leq \cdots \leq k_{t-1}$ and $i^r + r + j^s + k^t \leq n$.

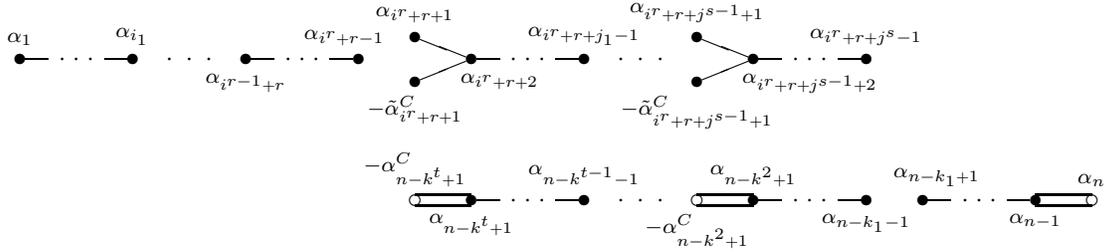
\unitlength0.75cm

\begin{picture}(19,4)

 \put(0,3){

\ACteil{0,0}
\Pure{2.5,0}
\ACteil{4,0}

\DCre{7,0}\ACteil{8,0}
\Pure{10.5,0}
\DCre{12,0}\ACteil{13,0}

\marke{0,0.25}{bc}{\alpha_1}
\marke{2,0.25}{bc}{\alpha_{i_1}}
\marke{4,-0.25}{tc}{\alpha_{i^{r \minus 1} \plus r}}
\marke{5.75,0.25}{bc}{\alpha_{i^r \plus r \minus 1}}

\marke{7,0.65}{bc}{\alpha_{i^r \plus r \plus 1}}
\marke{7,-0.65}{tc}{-\tilde\alpha^C_{i^r \plus r \plus 1}}
\marke{8.5,-0.25}{tc}{\alpha_{i^r \plus r \plus 2}}
\marke{10,0.25}{bc}{\alpha_{i^r \plus r \plus j_1 \minus 1}}

\marke{12,0.65}{bc}{\alpha_{i^r \plus r \plus j^{s-1} \plus 1}}
\marke{12,-0.65}{tc}{-\tilde\alpha^C_{i^r \plus r \plus j^{s-1} \plus 1}}
\marke{14,-0.25}{tc}{\alpha_{i^r \plus r \plus j^{s-1} \plus 2}}
\marke{15,0.25}{bc}{\alpha_{i^r \plus r \plus j^s \minus 1}}

 }

 \put(7,0.5){

\Cre{0,0}\ACteil{1,0}
\Pure{3.5,0}
\Cre{5,0}\ACteil{6,0}
\ACteil{9,0}
\Cli{12,0}

\marke{0,0.25}{bc}{-\alpha^C_{n \minus k^t \plus 1}}
\marke{1,-0.25}{tc}{\alpha_{n \minus k^t \plus 1}}
\marke{3,0.25}{bc}{\alpha_{n \minus k^{t \minus 1} \minus 1}}

\marke{5,-0.25}{tc}{-\alpha^C_{n \minus k^2 \plus 1}}
\marke{6,0.25}{bc}{\alpha_{n \minus k^2 \plus 1}}
\marke{8,-0.25}{tc}{\alpha_{n \minus k_1 \minus 1}}

\marke{9.25,0.25}{bc}{\alpha_{n \minus k_1 \plus 1}}
\marke{11,-0.25}{tc}{\alpha_{n \minus 1}}
\marke{12,0.25}{bc}{\alpha_n}

 }

\end{picture}

Closed root subsystems are those with $s = 0$, i.e., without $D$-factors.

\item[$(D_n)$]

Classes:\abs 
$
A_{i_1} \oplus \cdots \oplus A_{i_r} \oplus D_{j_1} \oplus \cdots \oplus D_{j_s}
$,
where $r,s \geq 0$, $i_1 \leq \cdots \leq i_r$, $j_1 \leq \cdots \leq j_s$ and $i^r + r + j^s \leq n$.

Bases:\abs\abs

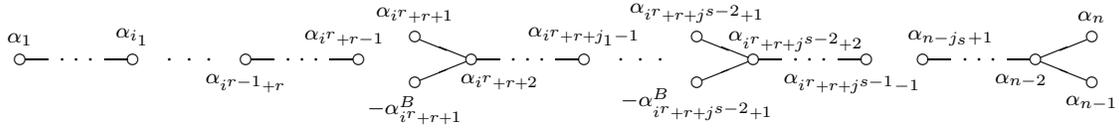
\unitlength0.75cm

\begin{picture}(15,2)

 \put(0,1){
\Ateil{0,0}
\Pure{2.5,0}
\Ateil{4,0}

\Dre{7,0}\Ateil{8,0}
\Pure{10.5,0}
\Dre{12,0}\Ateil{13,0}

\Ateil{16,0}
\Dli{19,0}

\marke{0,0.25}{bc}{\alpha_1}
\marke{2,0.25}{bc}{\alpha_{i_1}}
\marke{4,-0.25}{tc}{\alpha_{i^{r \minus 1} \plus r}}
\marke{5.75,0.25}{bc}{\alpha_{i^r \plus r \minus 1}}

\marke{7,0.65}{bc}{\alpha_{i^r \plus r \plus 1}}
\marke{7,-0.65}{tc}{-\alpha^B_{i^r \plus r \plus 1}}
\marke{8.5,-0.25}{tc}{\alpha_{i^r \plus r \plus 2}}
\marke{10,0.25}{bc}{\alpha_{i^r \plus r \plus j_1 \minus 1}}

\marke{12,0.65}{bc}{\alpha_{i^r \plus r \plus j^{s-2} \plus 1}}
\marke{12,-0.5}{tc}{-\alpha^B_{i^r \plus r \plus j^{s-2} \plus 1}}
\marke{13.75,0.15}{bc}{\alpha_{i^r \plus r \plus j^{s-2} \plus 2}}
\marke{14.75,-0.25}{tc}{\alpha_{i^r \plus r \plus j^{s \minus 1} \minus 1}}

\marke{16.5,0.25}{bc}{\alpha_{n \minus j_s + 1}}
\marke{17.75,-0.25}{tc}{\alpha_{n \minus 2}}
\marke{19,0.65}{bc}{\alpha_n}
\marke{19,-0.65}{tc}{\alpha_{n \minus 1}}
 }

\end{picture}

All root subsystems are closed.

\een

\esz

In case $k_l=1$, the factor $B_{k_l}$ or $C_{k_l}$ is isomorphic to $A_1$ and the
corresponding piece of the Dynkin diagram shrinks to  
\unitlength0.75cm
\begin{picture}(0.35,0.5)
\put(0.125,0.1){\circle*{0.2}}
\end{picture}
or
\unitlength0.75cm
\begin{picture}(0.35,0.5)
\put(0.125,0.1){\circle{0.2}}
\end{picture},
respectively. Similarly, in case $j_l=2$, the factor $D_{j_l}$ is
isomorphic to $A_1\oplus A_1$ and the corresponding piece of the Dynkin diagram 
shrinks to  
\unitlength0.75cm
\begin{picture}(0.35,0.5)
\put(0.125,0.3){\circle{0.2}}
\put(0.125,0){\circle{0.2}}
\end{picture}
or
\unitlength0.75cm
\begin{picture}(0.35,0.5)
\put(0.125,0.3){\circle*{0.2}}
\put(0.125,0){\circle*{0.2}}
\end{picture}.
Nevertheless, while being isomorphic, root subsystems obtained from one another by replacing a factor $B_1$ or $C_1$ by $A_1$, a factor $D_2$ by $A_1 \oplus A_1$ or a factor $D_3$ by $A_3$ are not conjugate under $W$.

\bbw

We assume that the elements of the base $\SSR$ of $\RS$ are numbered in such a way that the subset $\{\alpha_1 , \dots , \alpha_{n-1}\}$ is of class $A_{n-1}$. Starting from $\SSR$ and passing from lower to higher indices, we successively split off the necessary factors, always keeping the part $\tilde\SSR$ of $\SSR$ which contains $\alpha_n$. While $A$-factors can just be split off by removing one element from $\tilde\SSR$, before splitting off the other factors, one has to extend $\tilde\SSR$ by either the negative of its highest root or the negative of the dual of the highest root of $\tilde\SSR^\vee$. An elementary computation yields the following. Let $\tilde\SSR = \{\alpha_l , \dots , \alpha_n\}$.

\ben

\item For $B_n$, the highest root of $\tilde\SSR$ is $\alpha^B_l$ and the dual of the highest root of $\tilde\SSR^\vee$ is $\alpha^A_l$. 

\item For $C_n$, the highest root of $\tilde\SSR$ is $\alpha^C_l$ and the dual of the highest root of $\tilde\SSR^\vee$ is $\tilde\alpha^C_l$. 

\item For $D_n$, the highest root of $\tilde\SSR$ is $\alpha^D_l$. 

\een

Now, the assertion follows from the results in \cite{DyerLehrer} and \cite{Dynkin}.
\ebw

\begin{figure}

\centering

\unitlength1.75cm

\begin{picture}(8.5,8)

\put(0,1){

\put(0,7){
 \lri{0,0}{bc}{0}
 \whole{1,0}{bc}{A_1}
 }
\put(2,7){
 \lri{0,0}{bc}{0}
 \lri{1,0}{bc}{A_1}
 \whole{2,0}{bc}{A_2}
}
\put(5,7){
 \lri{0,0}{bc}{0}
 \lri{1,0}{bc}{A_1}
 \luri{1,0}{bc}{}
 \lri{2,0}{bc}{A_2}
 \lori{2,-0.5}{tc}{A_1 A_1}
 \whole{3,0}{bc}{A_3}
 }

\put(0,6){
 \lri{0,0}{bc}{0}
 \lri{1,0}{bc}{A_1}
 \luri{1,0}{bc}{}
 \lri{2,0}{bc}{A_2}
 \luri{2,0}{bc}{}
 \lori{2,-0.5}{tc}{A_1 A_1}
 \lri{2,-0.5}{tc}{}
 \lri{3,0}{bc}{A_3}
 \lori{3,-0.5}{tc}{A_1 A_2}
 \whole{4,0}{bc}{A_4}
 }


\put(5,4.75){
 \lori{0,0}{bc}{0}
 \lri{0,0}{bc}{}
 \lri{1,0.5}{bc}{B_1}
 \lri{1,0}{tc}{A_1}
 \lri{2,0}{tc}{D_2}
 \oluri{2,0.5}{bc}{B_1 B_1}
 \whole{3,0}{bc}{B_2}
 }


\put(0,3.5){
 \lori{0,0}{bc}{0}
 \lri{0,0}{bc}{}

 \lori{1,0.5}{br}{A_1}
 \lri{1,0.5}{}{}
 \luri{1,0.5}{}{}
 \lri{1,0}{bc}{B_1}
 \luri{1,0}{bc}{}
 
 \lri{2,1}{bc}{A_2}
 \lori{2,0}{tc}{A_1 B_1}
 \lori{2,0.5}{tc}{D_2}
 \lri{2,0.5}{tc}{}
 \luri{2,0.5}{tc}{}
 \olori{2,-0.5}{tc}{B_1 B_1}
 \olri{2,-0.5}{}{}
 
 \lurii{3,1}{bc}{D_3}
 \luri{3,0.5}{bc}{B_1 D_2}
 \lri{3,0}{bc}{B_2}
 \olori{3,-0.5}{tc}{B_1 B_1 B_1}
 
 \olri{4,0}{tl}{B_1 B_2}
 
 \whole{5,0}{cl}{B_3}

}


\put(3,1.5){
 \lori{0,0}{bc}{0}
 \lri{0,0}{bc}{}

 \lori{1,0.5}{br}{A_1}
 \lri{1,0.5}{}{}
 \luri{1,0.5}{}{}
 \lri{1,0}{bc}{C_1}
 \luri{1,0}{bc}{}
 

 \lri{2,1}{bc}{A_2}
 \lori{2,0}{tc}{A_1 C_1}
 \olori{2,0.5}{tc}{D_2}
 \olri{2,0.5}{tc}{}
 \oluri{2,0.5}{tc}{}
 \lori{2,-0.5}{tc}{C_1 C_1}
 \lri{2,-0.5}{}{}
 

 \olurii{3,1}{bc}{D_3}
 \oluri{3,0.5}{bc}{C_1 D_2}
 \lri{3,0}{bc}{C_2}
 \lori{3,-0.5}{tc}{C_1 C_1 C_1}
 
 
 \lri{4,0}{tl}{C_1 C_2}
 
 \whole{5,0}{cl}{C_3}

}


\put(0,0){
 \lri{0,0}{bc}{0}

 \lori{1,0}{br}{A_1}
 \lri{1,0}{}{}
 \luri{1,0}{}{}

 \lrii{2,0.5}{bc}{A_2}
 \lurri{2,0.5}{}{}
 \lorri{2,0}{tr}{A_1 A_1}
 \luri{2,0}{}{}
 \lorri{2,-0.5}{tc}{D_2}
 \lri{2,-0.5}{}{}

 \lri{3,-0.5}{tc}{A_1 D_2}

 \luri{4,0.5}{bc}{A_3}
 \lri{4,0}{tl}{D_3}
 \lori{4,-0.5}{tc}{D_2D_2}

 \whole{5,0}{cl}{D_4}

}

}

\end{picture}

\raggedright 

\caption{\label{Abb-HaDgr} Hasse diagrams of the sets of $W$-conjugacy classes of root subsystems for some low rank classical semisimple Lie algebras, cf.\ Proposition \rref{S-RT-klGr}. Closed root subsystems are represented by dots, non-closed ones by circles.} 

\end{figure}
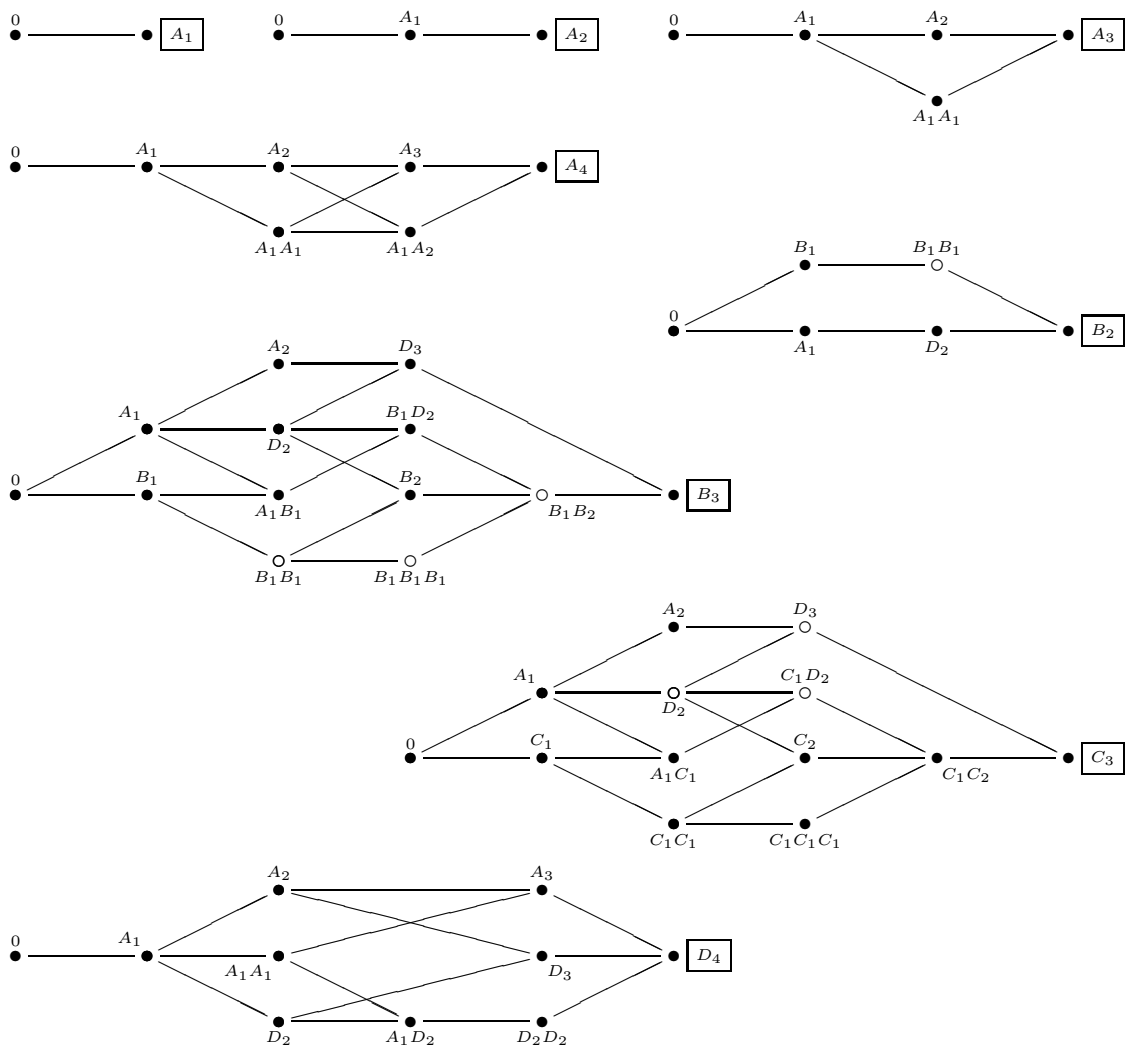

Figure \rref{Abb-HaDgr} shows the Hasse diagrams of the set $\RT$ of $W$-conjugacy classes of root subsystems for some examples. For each of the bases given in Proposition \rref{S-RT-klGr}, one can easily generate the corresponding root subsystem as a subset of $\RS$ by means of standard routines of the computer algebra system GAP \cite{GAP} with the package CHEVIE \cite{CHEVIE}. This is necessary for the computation of the coefficients $C^\rt_\hw$ and $D^\rt_\hw$, see Section \rref{A-Bsp}.

After having determined the set $\RT$ of conjugacy classes of root subsystems, in order to find the reflection type decomposition, from the subsets $T^\CC_\rt$, $\rt\in\RT$, one still has to single out those which are nonempty. In the case where $G$ is simply connected, this effort can be reduced by the following observation.

\bsz\label{S-RT}

If $G$ is simply connected, then $\RSS_x$ is closed for all $x\in T^\CC$. 

\esz

Thus, if $G$ is simply connected, one can restrict attention to closed root
subsystems. Of course, this does not make a difference unless $\mf g^\CC$
contains a factor $B_l$, $C_l$, $F_4$ or $G_2$.

\bbw

Let $\alpha,\beta\in\RSS_x$ such that $\alpha + \beta \in \RS$. Since $G$ is simply connected, we have $q_\alpha = p_\alpha = 1$ and $q_\beta = p_\beta = 1$. Therefore, Proposition \rref{S-Gx} yields $\mr e_\alpha(x) = \mr e_\beta(x) = 1$ and hence
$$
E_{\alpha + \beta}(x)
 = 
\mr e_{\alpha + \beta}(x) - 1
 =
\mr e_\alpha(x) \, \mr e_\beta(x) - 1
 =
0\,,
$$
i.e., $\alpha + \beta\in\RSS_x$.
\ebw

If $G$ is not simply connected, the argument fails, because here one has $E_\alpha = e_{\frac{q_\alpha}{p_\alpha} \, \alpha} - 1$ and there
is no simple relation between the ratios $\frac{q_\alpha}{p_\alpha}$ for the
elements of $\RSS$ and the corresponding ratios of their sums. This is illustrated by the following example.

\bbs\label{Bsp-RT}

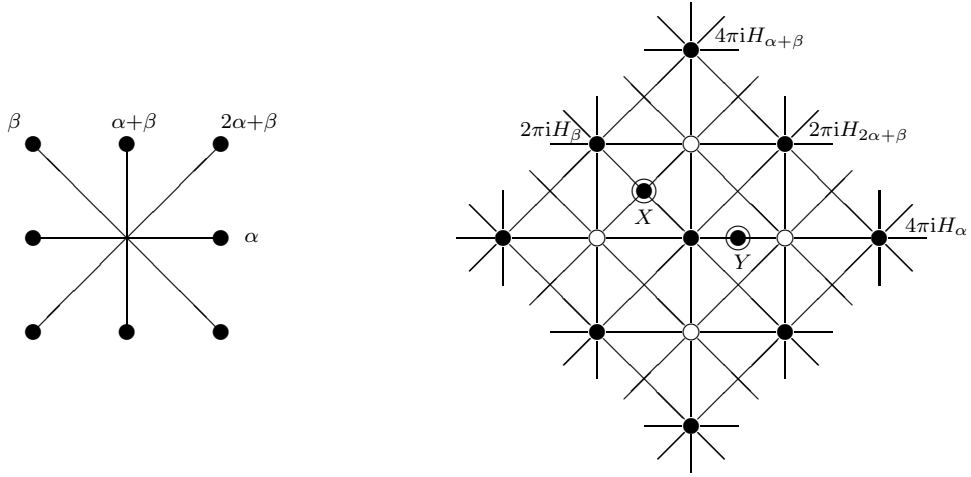
\begin{figure}

\unitlength1.25cm

\centering

\begin{picture}(12,6)
 \put(3,3){

\punkt{1,1}
\punkt{1,-1}
\punkt{-1,1}
\punkt{-1,-1}
\punkt{1,0}
\punkt{0,1}
\punkt{-1,0}
\punkt{0,-1}

\linie{0,0}{-1,0}{1}
\linie{0,0}{1,0}{1}
\linie{0,0}{0,1}{1}
\linie{0,0}{0,-1}{1}
\linie{0,0}{1,1}{1}
\linie{0,0}{1,-1}{1}
\linie{0,0}{-1,1}{1}
\linie{0,0}{-1,-1}{1}

\marke{1.25,0}{cl}{\alpha}
\marke{1,1.125}{bl}{2\alpha+\beta}
\marke{0.075,1.125}{bc}{\alpha+\beta}
\marke{-1.125,1.125}{br}{\beta}
 }

 \put(9,3){

\punkt{0,0}
\punkt{2,0}
\punkt{-2,0}
\punkt{0,2}
\punkt{0,-2}
\punkt{1,1}
\punkt{1,-1}
\punkt{-1,1}
\punkt{-1,-1}

\kreis{1,0}
\kreis{0,1}
\kreis{-1,0}
\kreis{0,-1}





\hll{-2,0}
\lr{-2,0}
\lr{-1,0}
\lr{0,0}
\lr{1,0}
\hlr{2,0}

\hll{-1,1}
\lr{-1,1}
\lr{0,1}
\hlr{1,1}

\hll{-1,-1}
\lr{-1,-1}
\lr{0,-1}
\hlr{1,-1}

\hll{0,2}
\hlr{0,2}

\hll{0,-2}
\hlr{0,-2}


\hlu{-2,0}
\hlo{-2,0}

\hlu{-1,-1}
\lo{-1,-1}
\lo{-1,0}
\hlo{-1,1}

\hlu{0,-2}
\lo{0,-2}
\lo{0,-1}
\lo{0,0}
\lo{0,1}
\hlo{0,2}

\hlu{1,-1}
\lo{1,-1}
\lo{1,0}
\hlo{1,1}

\hlu{2,0}
\hlo{2,0}


\vllu{-2,0}

\vllu{-1,-1}
\dvllu{-1,0}
\lro{-2,0}

\vllu{0,-2}
\dvllu{0,-1}
\lro{-1,-1}
\lro{-1,0}
\lro{-1,1}

\lro{0,-2}
\lro{0,-1}
\lro{0,0}
\dvlro{0,1}
\vlro{0,2}

\lro{1,-1}
\dvlro{1,0}
\vlro{1,1}

\vlro{2,0}


\vllo{-2,0}

\lru{-2,0}
\dvllo{-1,0}
\vllo{-1,1}

\lru{-1,-1}
\lru{-1,0}
\lru{-1,1}
\dvllo{0,1}
\vllo{0,2}

\vlru{0,-2}
\dvlru{0,-1}
\lru{0,0}
\lru{0,1}
\lru{0,2}

\vlru{1,-1}
\dvlru{1,0}
\lru{1,1}

\vlru{2,0}


\punkt{-0.5,0.5}
\punkt{0.5,0}
\kkreis{-0.5,0.5}
\kkreis{0.5,0}

\marke{-0.5,0.3}{tc}{X}
\marke{0.55,-0.175}{tc}{Y}

\marke{2.275,0.125}{cl}{4\pi \mr i H_\alpha}
\marke{1.25,1.125}{cl}{2\pi \mr i H_{2 \alpha+\beta}}
\marke{0.25,2.125}{cl}{4\pi \mr i H_{\alpha+\beta}}
\marke{-1.15,1.125}{cr}{2\pi \mr i H_{\beta}}
 }

\end{picture}

\caption{\label{Abb-Bsp-RT} Root system $\RS$ (a) and Cartan subalgebra $\mf t$
(b) of Example \rref{Bsp-RT}. In (b), the elements of $\ker(\exp_{\tilde G})$
and of $\ker(\exp_G) \setminus \ker(\exp_{\tilde G})$ are represented by dots
and circles, respectively.} 

\end{figure}

Consider $G = \SO(5)$ and its (twofold) universal covering group $\tilde G =
\Sp(2) \cong \Spin(5)$. Every base of the root system $\RS$ consists of a short
root $\alpha$ and a long root $\beta$ and  
$$
\RS = \{\pm\alpha, \pm\beta, \pm(\alpha+\beta), \pm(2\alpha+\beta)\}\,,
$$
see Figure \rref{Abb-Bsp-RT}(a). The Killing form can be chosen so that $\kf(\alpha,\alpha) = 1 $. Then,
$$
\kf(\alpha,\beta) = -1
 \,,~~~~
\kf(\beta,\beta) = 2
 \,,~~~~
\kf(\alpha+\beta,\alpha+\beta) = 1
 \,,~~~~
\kf(2\alpha+\beta,2\alpha+\beta) = 2\,.
$$
The elements $2\pi\mr i \frac{2 H_\gamma}{\kf(\gamma,\gamma)}$, $\gamma\in\RS$, of $\mf t$ are thus given, respectively, by 
$$
\pm4\pi\mr i H_\alpha
 \,,~~~~
\pm2\pi\mr i H_\beta
 \,,~~~~
\pm4\pi\mr i H_{\alpha + \beta}
 \,,~~~~
\pm2\pi\mr i H_{2\alpha+\beta}
$$
and $\ker(\exp_{\tilde G})$ is generated by $4\pi\mr i H_\alpha$ and $2\pi\mr i H_\beta$, whereas $\ker(\exp_G)$ is generated by $2\pi\mr i H_\alpha$ and $2\pi\mr i H_\beta$, see Figure \rref{Abb-Bsp-RT}(b). Consider the subsets of long roots and of short roots,
$$
\RSS_l = \{\pm\beta,\pm(2\alpha+\beta)\}
 \,,~~~~~~
\RSS_s = \{\pm\alpha,\pm(\alpha+\beta)\}\,.
$$
Both of them are root subsystems. They represent the classes $C_1 \oplus C_1$ and $D_2$, respectively, of Proposition \rref{S-RT-klGr}. While $\RSS_l$ is closed, $\RSS_s$ is not. 
Hence, only $\RSS_l$ may represent a reflection type for $\tilde G$. It does, indeed: for $X = \pi\mr i H_\beta$, from Figure \rref{Abb-Bsp-RT}(b) we read off
 \ala{
\sigma_\alpha(X) - X & = 2 \pi \mr i H_\alpha
 \,, &
\sigma_\beta(X) - X & = - 2 \pi \mr i H_\beta 
 \,, \\
\sigma_{\alpha+\beta}(X) - X & = - 2 \pi \mr i H_{\alpha+\beta}
 \,, &
\sigma_{2\alpha+\beta}(X) - X & = 0 \,.
 }
Since $- 2 \pi \mr i H_\beta$ belongs to $\ker(\exp_{\tilde G})$, whereas $2 \pi \mr i H_\alpha$ and $- 2 \pi \mr i H_{\alpha+\beta}$ do not, we obtain $\RSS_{\exp_{\tilde G} X} = \RSS_l$. For $G$, things are different: for $Y = \pi\mr i H_\alpha$, from Figure \rref{Abb-Bsp-RT}(b) we read off
 \ala{
\sigma_\alpha(Y) - Y & = - 2 \pi \mr i H_\alpha
 \,, &
\sigma_\beta(Y) - Y & = \pi \mr i H_\beta
 \,, \\
\sigma_{\alpha+\beta}(Y) - Y & = 0
 \,, &
\sigma_{2\alpha+\beta}(Y) - Y & = - \pi \mr i H_{2\alpha+\beta} \,.
 }
Since $- 2 \pi \mr i H_\alpha$ belongs to $\ker(\exp_G)$, whereas $\pi \mr i H_\beta$ and $- \pi \mr i H_{2\alpha+\beta}$ do not, $\RSS_{\exp_G Y} = \RSS_s$. Thus, $\RSS_s$ represents a reflection type for $G$.
\qeb

\ebs

\bbm\label{Bem-RT-HO}~

By analogy, one can define reflection types for the action of $W$ on $\mf t^\CC$ induced from the adjoint representation of $G$ on $\mf g$. Let us refer to these reflection types as infinitesimal reflection types. Since $\exp$ is locally a diffeomorphism, it is clear that every infinitesimal reflection type is a reflection type of the action of $W$ on $T^\CC$. Conversely, since $\eta\in\mf t^\CC$ is invariant under a reflection $\sigma_\alpha$ iff $\alpha(\eta) = 0$, the reflection type defined by a closed root subsystem $\RSS$ of $\RS$ is infinitesimal iff there is no larger root subsystem $\RSS'\supset\RSS$ of $\RS$ such that $\text{span}_\CC\RSS = \text{span}_\CC\RSS'$. 
This means that there is no larger root subsystem of the same rank. Thus, for $A_n$, all reflection types are infinitesimal, whereas all the other classical series admit reflection types which are not infinitesimal, e.g., $C_{n-1} \oplus C_1$ for $C_n$ with $n\geq 2$, $D_n$ for $B_n$ with $n\geq 3$ and $D_2\oplus D_{n-2}$ for $D_n$ with $n\geq 4$.
In particular, the reflection types discussed in Example \rref{Bsp-RT} are not infinitesimal, because they have maximal rank. 
\qeb

\ebm


\section{Examples}
\label{A-ex}\label{A-Bsp}


As a test, we computed the coefficients $\tilde C^\rt_\hw$ and $\tilde D^\rt_\mu$ for all conjugacy classes of closed root subsystems for the classical Lie algebras up to rank $4$ using the computer algebra system GAP \cite{GAP} with the package CHEVIE \cite{CHEVIE} and Mathematica \cite{Mathematica} with the package LieART \cite{LieART}. According to Proposition \rref{S-RT}, the closed root subsystems yield all candidates for the reflection types of the corresponding simply connected compact Lie groups. We proceeded as follows. Let $\rt$ be one of the conjugacy classes of closed root subsystems from Proposition \rref{S-RT-klGr}. To start with, we use GAP and CHEVIE to generate a list {\tt COMPLEMENT} of the elements of $\RS \setminus \RSS$ from the corresponding base given in that Proposition. Using {\tt COMPLEMENT}, we compile a list {\tt WEYL2} containing one representative of each $W_\RSS$-coset in $W$. Then, we generate a list {\tt SUBSETSUM1} of all pairs $\big(\eta,\vfh\RSS(\eta)\big)$, where $\eta = \hw_\CSS$ for some subset of $\CSS \subset \RS \setminus \RSS$, as follows. Starting with the single pair $(\eta,V) = (0,1)$, in step $n$, for each pair $(\eta,V)$ of the list from step $n-1$, we append to this list a new pair $(\eta + {\tt COMPLEMENT}[n],-V)$. From the enlarged list so produced we obtain the list of step $n$ by collecting pairs with the same $\eta$ and taking the sum over all the associated $V$'s. This procedure circumvents the generation of all subsets of $\RS \setminus \RSS$. Next, we generate a list {\tt SUBSETSUM2} of all pairs
$\big(\eta,\tvfh\RSS(\eta)\big)$, where $\eta = w'^{-1}\hw_\CSS$ for some subset
$\CSS \subset \RS \setminus \RSS$ and the chosen representative $w'$ of some
right $W_\RSS$-coset in $W$. Starting with {\tt SUBSETSUM1}, in step $n$, we take the $n$-th element $w'$ of {\tt WEYL2} and successively check for each $(\eta,V)$ in the list of step $n-1$ whether $w'^{-1}(\eta) = \eta'$ for some pair $(\eta',V')$. If so, we replace $V'$ by $V' + V$ in this pair. If not, we add the pair $\big(w'^{-1}(\eta),V\big)$ to the list. Finally, we produce a list {\tt IRREP} of all $\hw\in\HW$ satisfying
\beq\label{G-Normabschaetzung}
\|\hw + \delta\| \leq M + \|\delta\|\,,
\eeq
where $M$ is the maximum of the norms $\|\eta\|$ over all pairs $(\eta,V)$ in
{\tt SUBSETSUM1}. This list contains all elements of $\HW^\rt$, because for
every subset $\CSS \subset \RS \setminus \RSS$, the relation $w'\big(w''(\hw +
\delta) - \delta\big) = \lambda_\CSS$ implies \eqref{G-Normabschaetzung}. Now, using {\tt
SUBSETSUM2}, for every member $\hw$ of {\tt IRREP}, we compute the quotient
$\tilde C^\rt_\hw/N_\hw$ of the reduced coefficient $\tilde C^\rt_\hw$ by means
of Formula \eqref{G-Cl-red}. By omitting members whose coefficient vanishes, we
finally obtain a list {\tt COEFF} containing all pairs $(\hw,\tilde
C^\rt_\hw/N_\hw)$ with $\hw\in\HW^\rt$.

The computation of the reduced coefficients $\tilde D^\rt_\mu$ is
straighforward. First, using the routine DominantWeightSystem of LieART, we
generate a list {\tt MULT} of triples $\big(\lambda,\mu,m_\lambda(\mu)\big)$,
where $\hw\in\HW^\rt$ and $\mu\in\HWS(\lambda)$. From this, we extract a list of
the elements $\mu$ of $\HWS^\rt$ , i.e., the dominant weights of all the irreps
contained in $\HW^\rt$. Using {\tt COEFF}, for each member of this list, we
compute the reduced coefficient $\tilde D^\rt_\mu$.  

Tables with the results are given in the appendix. For lack of space, we omit the results for $B_4$ and $C_4$.

\bbm

We compare our results with those of \cite{Hue:Bedlewo}, where the case of
$G=\SU(n+1)$ was discussed. Let $\hw_1,\dots,\hw_n$ denote the fundamental
weights of $A_n$. Using the mapping 
$$
\vp
 := 
 \left(
-\chi^\CC_{\hw_1},\chi^\CC_{\hw_2},\dots,(-1)^n\chi^\CC_{\hw_n},(-1)^{n+1}
 \right)
 : 
T^\CC \to \CC^{n+1}\,,
$$
embedding $T^\CC$ into the space of normalized complex polynomials of degree
$n+1$ with constant coefficient being equal to $(-1)^{n+1}$, it was shown there that
for $k=1,\dots,n$, the subset $T^\CC_{(A_k)}$ of $T^\CC_{(A_{k-1})}$ (in our
notation and with $T^\CC_{(A_0)} = T^\CC$ understood) is defined by a relation
of the type $D \circ \vp = 0$, where $D$ is the discriminant of a certain
polynomial of degree $n+2-k$. For example, denoting the discriminant of the polynomial $a_0 z^n + a_1 z^{n-1} + \cdots + a_n$ by $D_n(a_0,\dots,a_n)$, the relations are
 \al{\label{G-Rel-Joh-11}
D_2(1,-\chi_{\hw_1},1) & = 0
 & &
\text{defining } T^\CC_{(A_1)} \subset T^\CC\,,
\\ 
\intertext{for $n=1$,}
\label{G-Rel-Joh-21}
D_3(1,-\chi_{\hw_1},\chi_{\hw_2},-1) & = 0
 & &
\text{defining } T^\CC_{(A_1)} \subset T^\CC\,,
\\ \label{G-Rel-Joh-22}
D_2(3,-2\chi_{\hw_1},\chi_{\hw_2}) & = 0
 & &
\text{defining } T^\CC_{(A_2)} \equiv T^\CC_{A_2} \subset T^\CC_{(A_1)}
\\
\intertext{for $n=2$ and}
\label{G-Rel-Joh-31}
D_4(1,-\chi_{\hw_1},\chi_{\hw_2},-\chi_{\hw_3},1) & = 0
 & &
\text{defining } T^\CC_{(A_1)} \subset T^\CC\,,
\\ \label{G-Rel-Joh-32}
D_3(4,-3\chi_{\hw_1},2\chi_{\hw_2},-\chi_{\hw_3}) & = 0
 & &
\text{defining } T^\CC_{(A_2)} \subset T^\CC_{(A_1)}\,,
\\ \label{G-Rel-Joh-33}
D_2(12,-6\chi_{\hw_1},2\chi_{\hw_2}) & = 0
 & &
\text{defining } T^\CC_{(A_3)} \equiv T^\CC_{A_3} \subset T^\CC_{(A_2)}.
 }
for $n=3$. A direct computation shows that \eqref{G-Rel-Joh-11}, \eqref{G-Rel-Joh-21} and \eqref{G-Rel-Joh-31} coincide with the relation $\rel0 = 0$ in the respective case and that $\rel{A_1} = 0$ for $n=2$ coincides with 
$$
D_2(3,-2\chi_{\hw_1},\chi_{\hw_2}) D_2(3,-2\chi_{\hw_2},\chi_{\hw_1}) = 0\,,
$$
which is equivalent to \eqref{G-Rel-Joh-22}. The other relations are not directly comparable, because they have different meanings. For example, while the relation \eqref{G-Rel-Joh-32} means that solutions in $T^\CC_{(A_1)}$ {\em do} belong to the subset $T^\CC_{(A_2)}$, the relation $\rel{A_1}=0$ means that solutions in $T^\CC_{(A_1)}$ {\em do not} belong to $T^\CC_{A_1}$ and thus belong to the rest, which is $T^\CC_{(A_2)} \cup T^\CC_{(A_1\oplus A_1)}$.
\qeb

\ebm

\bbm\label{Bem-SU(n)}\label{Bem-RT-HO}~

It remains to analyze the precise difference between the decompositions by orbit
types and by reflection types. We do not discuss this here. Let us just state
that a direct inspection yields that in the cases $G=\SU(n+1)$ and $G=\Sp(n)$,
reflection types correspond bijectively to orbit types and, therefore, the
reflection type subsets coincide with the orbit type subsets. Indeed, the
stabilizers under the adjoint action of $G$ are given by the Howe
subgroups\footnote{subgroups $H\subseteq G$ satisfying $\mr C_G^2(H) = H$} of
$G$ containing a maximal torus. By \cite{HSG}, these subgroups are given as
follows. For $G=\SU(n+1)$, they are labelled by a set of positive integers
$\{k_1,\dots,k_r\}$ such that $\sum_{i=1}^r k_i = n+1$. The class labelled by
such a set is represented by the subgroup 
$$
H
 = 
 \left\{
\diag(a_1,\dots,a_r) : a_i \in \mr U(k_i) \,, ~ \prod_{i=1}^r \det(a_i) = 1 
 \right\} \,,
$$
acting on $\CC^{k_1} \oplus \cdots \oplus \CC^{k_r}$. 
For $G=\Sp(n)$, they are labelled by two sets of positive integers $\{k_1,\dots,k_r\}$, $\{l_1,\dots,l_s\}$ such that $\sum_{i=1}^r k_i + \sum_{i=1}^s l_i = n$ and the class labelled by such a pair of sets is represented by the subgroup
$$
H
 = 
 \left\{
\diag(a_1,\dots,a_r,b_1,\dots,b_s) : a_i \in \mr U(k_i) \,, b_i \in \Sp(l_i)
 \right\}\,,
$$
acting on $\HH^{k_1} \oplus \cdots \oplus \HH^{k_r} \oplus \HH^{l_1} \oplus
\cdots \oplus \HH^{l_s}$. To $H$ there corresponds the subgroup 
$$
W_H := (H\cap\mr N_G(T))/T
$$
of $W$, acting on $\mf t^\CC$. In the case $G = \SU(n+1)$, we choose $T$ to consist of the diagonal matrices in $G$. Then, $\mf t^\CC$ consists of the diagonal matrices in $\sl(n+1,\CC)$ and $W_H$ consists of the elements $\pi_1 \times \cdots \times \pi_r$, where $\pi_i$ is a permutation of $k_i$ elements and $\pi_1$ acts on the first $k_1$ entries, $\pi_2$ acts on the entries number $k_1+1$ up to $k_1+k_2$ etc.
If we further choose the elements $\alpha_i$, $i=1,\dots,n$, of the base $\PS_0$
to be given by the difference between the $(i+1)$th and the $i$th entry, then
$W_H$ is generated by the reflections associated with the simple roots
$\alpha_{k(i-1)+1},\dots,\alpha_{k(i)}$ for all $i$ such that $k_i\geq 2$. Thus,
$W_H$ represents the reflection type labelled by the regular semisimple
subalgebra $A_{k_1-1} \oplus \cdots \oplus A_{k_r-1}$, where factors $A_0 \equiv
0$ can be omitted.  

In the case $G = \Sp(n)$, we write quaternionic $(n\times n)$-matrices as complex $(2n\times 2n)$-matrices by replacing the quaternion $q$ by the $(2\times 2)$-matrix 
$$
\bbma \ol{q_1} & - \ol{q_2} \\ q_2 & q_1 \ebma\,,
$$
where $q = q_1 + q_2 j$ with $q_1,q_2\in\CC$ and $j$ denoting the second quaternionic unit.
Then, $\mf g^\CC = \sp(n,\CC)$. In analogy with the case $G=\SU(n+1)$, by choosing a concrete representation of $T$ and a concrete base $\PS_0$, one can check that $W_H$ represents the reflection type labelled by $A_{k_1-1} \oplus \cdots \oplus A_{k_r-1} \oplus C_{l_1} \oplus \cdots \oplus C_{l_s}$, where factors $A_0 \equiv 0$ can be omitted again.
\qeb

\ebm

\section*{Acknowledgments}

The authors acknowledge discussions with Erik Fuchs, Florian F\"urstenberg,
Michael Schellenberger, Christian Fleischhack and Rainer Matthes.

\newpage

\begin{appendix}

\section{Tables}

In the following tables, dominant weights $\hw\in\HW$ are given by their
Dynkin labels $\frac{2\kf(\hw,\alpha_i)}{\kf(\alpha_i,\alpha_i)}$.
For each closed root subsystem, the first column
contains the reduced coefficient $\tilde C^\rt_\hw/N_\hw$ and the second column
contains the reduced coefficient $\tilde D^\rt_\hw$. 
\bigskip

\renewcommand{\tabcolsep}{2.5pt}
\begin{tabular}{ll}

\begin{tabular}[t]{l}

$\SU(2)$  ($A_1$)
\\
$
\tiny
\renewcommand{\arraycolsep}{3pt}
\begin{array}{|c||c|c|}
\hline
\text{\footnotesize $\lambda$}
 & 
\multicolumn{2}{c|}{0}
\\ \hline
 0 & 3 & 2 
\\ \hline
 2 & -1 & -1 
\\ \hline
\end{array}
$
\bigskip
\\
\\

$\SU(3)$  ($A_2$)
\\
$
\tiny
\renewcommand{\arraycolsep}{3pt}
\begin{array}{|c||c|c||c|c|}
\hline
 \lambda
 & 
\multicolumn{2}{c||}{0}
 &
\multicolumn{2}{|c|}{A_1}
\\ \hline
 00 & 15 & 6 & 20 & 12
\\ \hline
 03 & 3 & 2 & 1 & 1
\\ \hline
 11 & -6 & -2 & -5 & -3
\\ \hline
 22 & -1 & -1 & &
\\ \hline
 30 & 3 & 2 & 1 & 1
\\ \hline
\end{array}
$
\bigskip
\\
\\

$\SU(4)$  ($A_3$)
\\
$
\tiny
\renewcommand{\arraycolsep}{2pt}
\begin{array}{|c||c|c||c|c||c|c||c|c|}
\hline
 \lambda
 & 
\multicolumn{2}{c||}{0}
 &
\multicolumn{2}{c||}{A_1}
 &
\multicolumn{2}{c||}{A_1 \oplus A_1}
 &
\multicolumn{2}{c|}{A_2}
\\ \hline
000 & 105 & 24 & 350 & 120 & 105 & 54 & 50 & 32
\\ \hline
004 & -15 & -6 & -20 & -12 & & & -1 & -1 
\\ \hline
012 & 27 & 4 & 54 & 14 &  & & 3 & 2 
\\ \hline
020 & -6 & 4 & -14 & 16 & 21 & 16 & -8 & -4
\\ \hline
032 & -3 & -2 & -1 & -1 & & & &  
\\ \hline
040 & 9 & 4 & 6 & 4 & 1 & 1 & & 
\\ \hline
101 & -45 & -6 & -120 & -27 & -35 & -12 & -6 & -4
\\ \hline
113 & 6 & 2 & 5 & 3 & & & & 
\\ \hline
121 & -12 & -2 & -16 & -5 & -3 & -2 & & 
\\ \hline
202 & -9 & -4 & -15 & -12 & 5 & 3 & & 
\\ \hline
210 & 27 & 4 & 54 & 14 & & & 3 & 2
\\ \hline
222 & 1 & 1 & & & & & & 
\\ \hline
230 & -3 & -2 & -1 & -1 & & & & 
\\ \hline
303 & -3 & -2 & -1 & -1 & & & & 
\\ \hline
311 & 6 & 2 & 5 & 3 & & & & 
\\ \hline
400 & -15 & -6 & -20 & -12 & & & -1 & -1
\\ \hline
\end{array}
$
\bigskip
\\
\\
\\
\\
\\
\\

$\Sp(2)$ ($C_2$)
\\
$
\tiny
\renewcommand{\arraycolsep}{2pt}
\begin{array}{|c||c|c||c|c||c|c||c|c|}
\hline
\lambda
 & 
\multicolumn{2}{c||}{0}
 & 
\multicolumn{2}{c||}{C_1}
 & 
\multicolumn{2}{c||}{A_1}
 & 
\multicolumn{2}{c|}{C_1 \oplus C_1}
\\ \hline
00 & 21 & 8 & 21 & 12 & 28 & 16 & 5 & 4
\\ \hline
01 & -3 & -2 & -7 & -4 & 6 & -1 & -3 & -2
\\ \hline
02 & -3 & & -4 & -2 & 3 & 4 &   & 
\\ \hline
03 & -3 & -2 & & & -1 & -1 &   &
\\ \hline
20 & -6 & -2 & & & -14 & -8 & 1 & 1
\\ \hline
21 & 6 & 2 & 3 & 2 & 2 & 1 &   &
\\ \hline
22 & 1 & 1 & & & & & &
\\ \hline
40 & -3 & -2 & -1 & -1 &   &   &   &
\\ \hline
\end{array}
$

\bigskip

\end{tabular}

&

\begin{tabular}[t]{l}

$\SU(5)$  ($A_4$)
\\
$
\tiny
\renewcommand{\arraycolsep}{2pt}
\begin{array}{|c||c|c||c|c||c|c||c|c||c|c||c|c|}
\hline
 \lambda
 & 
\multicolumn{2}{c||}{0}
 & 
\multicolumn{2}{c||}{A_1}
 & 
\multicolumn{2}{c||}{A_1 \oplus A_1}
 & 
\multicolumn{2}{c||}{A_2}
 & 
\multicolumn{2}{c||}{A_1 \oplus A_2}
 & 
\multicolumn{2}{c|}{A_3}
\\ \hline
 0000 & 945 & 120 & 6300 & 1200 & 6720 & 1920 & 1899 & 740 & 1520 & 880 & 105 & 80 
\\ \hline
 0005 & 105 & 24 & 350 & 120 & 105 & 54 & 50 & 32 &   &   & 1 & 1 
\\ \hline
 0013 & -180 & -12 & -750 & -84 & -230 & -39 & -132 & -40 & 10 & 4 & -3 & -2 
\\ \hline
 0021 & 45 & -12 & 210 & -96 & -250 & -189 & 198 & 32 & 2 & -12 & 3 & 1 
\\ \hline
 0042 & -15 & -6 & -20 & -12 &   &   & -1 & -1 	&   &   &   &   
\\ \hline
 0050 & 45 & 12 & 75 & 30 & 20 & 12 & 3 & 2 & 1 & 1 &   &   
\\ \hline
 0102 & 270 & 12 & 1350 & 102 & 945 & 120 & 165 & 40 & -45 & -2 & 6 & 4 
\\ \hline
 0110 & -75 & 12 & -400 & 108 & 225 & 195 & -285 & 16 & 168 & 80 & -21 & -7 
\\ \hline
 0123 & 27 & 4 & 54 & 14 &   & 0 & 3 & 2 &   &   &   &   
\\ \hline
 0131 & -54 & -4 & -135 & -18 & -45 & -12 & -6 & -2 & -3 & -2 &   &   
\\ \hline
 0204 & -6 & 4 & -14 & 16 & 21 & 16 & -8 & -4 &   &   &   &   
\\ \hline
 0212 & -27 & -8 & -81 & -44 & -18 & -20 & 15 & 4 & 3 & 1 &   &   
\\ \hline
 0220 & 111 & 8 & 370 & 48 & 90 & 0 & 6 & 4 & 12 & 8 &   &   
\\ \hline
 0301 & -18 & 4 & -66 & 26 & 25 & 36 & -9 & 6 & -7 & 5 &   &   
\\ \hline
 0322 & -3 & -2 & -1 & -1 &   &   &   &   &   &   &   &   
\\ \hline
 0330 & 9 & 4 & 6 & 4 & 1 & 1 &   &   &   &   &   &   
\\ \hline
 0403 & 9 & 4 & 6 & 4 & 1 & 1 &   &   &   &   &   &   
\\ \hline
 0411 & -18 & -4 & -21 & -8 & -5 & -3 &   &   &   &   &   &   
\\ \hline
 0500 & 45 & 12 & 75 & 30 & 20 & 12 & 3 & 2 & 1 & 1 &   &   
\\ \hline
 1001 & -420 & -24 & -2450 & -228 & -2520 & -360 & -382 & -106 & -411 & -141 &   & -4 
\\ \hline
 1014 & -45 & -6 & -120 & -27 & -35 & -12 & -6 & -4 &   &   &   &   
\\ \hline
 1022 & 81 & 4 & 270 & 24 & 135 & 27 & 9 & 0 & 6 & 4 &   &   
\\ \hline
 1030 & -18 & 4 & -66 & 26 & 25 & 36 & -9 & 6 & -7 & 5 &   &   
\\ \hline
 1103 & 72 & 8 & 264 & 52 & -10 & 12 & 33 & 20 & -13 & -5 &   &   
\\ \hline
 1111 & -144 & -2 & -600 & -15 & -220 & 6 & -108 & -26 & -22 & -15 &   &   
\\ \hline
 1132 & 6 & 2 & 5 & 3 &   &   &   &   &   &   &   &   
\\ \hline
 1140 & -18 & -4 & -21 & -8 & -5 & -3 &   &   &   &   &   &   
\\ \hline
 1200 & 45 & -12 & 210 & -96 & -250 & -189 & 198 & 32 & 2 & -12 & 3 & 1 
\\ \hline
 1213 & -12 & -2 & -16 & -5 & -3 & -2 &   &   &   &   &   &   
\\ \hline
 1221 & 24 & 2 & 44 & 7 & 18 & 6 &   &   &   &   &   &   
\\ \hline
 1302 & 18 & 4 & 39 & 16 & 2 & 7 &   &   &   &   &   &   
\\ \hline
 1310 & -54 & -4 & -135 & -18 & -45 & -12 & -6 & -2 & -3 & -2 &   &   
\\ \hline
 2002 & -90 & -18 & -420 & -144 & -175 & -144 & -6 & -28 & 128 & 72 &   &   
\\ \hline
 2010 & 270 & 12 & 1350 & 102 & 945 & 120 & 165 & 40 & -45 & -2 & 6 & 4 
\\ \hline
 2023 & -9 & -4 & -15 & -12 & 5 & 3 &   &   &   &   &   &   
\\ \hline
 2031 & 18 & 4 & 39 & 16 & 2 & 7 &   &   &   &   &   &   
\\ \hline
 2104 & 27 & 4 & 54 & 14 &   & 0 & 3 & 2 &   &   &   &   
\\ \hline
 2112 & -36 & 0 & -96 & 0 & -37 & -12 & -9 & -4 &   &   &   &   
\\ \hline
 2120 & -27 & -8 & -81 & -44 & -18 & -20 & 15 & 4 & 3 & 1 &   &   
\\ \hline
 2201 & 81 & 4 & 270 & 24 & 135 & 27 & 9 & 0 & 6 & 4 &   &   
\\ \hline
 2222 & 1 & 1 &   &   &   &   &   &   &   &   &   &   
\\ \hline
 2230 & -3 & -2 & -1 & -1 &   &   &   &   &   &   &   &   
\\ \hline
 2303 & -3 & -2 & -1 & -1 &   &   &   &   &   &   &   &   
\\ \hline
 2311 & 6 & 2 & 5 & 3 &   &   &   &   &   &   &   &   
\\ \hline
 2400 & -15 & -6 & -20 & -12 &   &   & -1 & -1 &   &   &   &   
\\ \hline
 3003 & -36 & -12 & -114 & -66 & 56 & 24 & -6 & -10 & 1 & 1 &   &   
\\ \hline
 3011 & 72 & 8 & 264 & 52 & -10 & 12 & 33 & 20 & -13 & -5 &   &   
\\ \hline
 3032 & -3 & -2 & -1 & -1 &   &   &   &   &   &   &   &   
\\ \hline
 3040 & 9 & 4 & 6 & 4 & 1 & 1 &   &   &   &   &   &   
\\ \hline
 3100 & -180 & -12 & -750 & -84 & -230 & -39 & -132 & -40 & 10 & 4 & -3 & -2 
\\ \hline
 3113 & 6 & 2 & 5 & 3 &   &   &   &   &   &   &   &   
\\ \hline
 3121 & -12 & -2 & -16 & -5 & -3 & -2 &   &   &   &   &   &   
\\ \hline
 3202 & -9 & -4 & -15 & -12 & 5 & 3 &   &   &   &   &   &   
\\ \hline
 3210 & 27 & 4 & 54 & 14 &   & 0 & 3 & 2 &   &   &   &   
\\ \hline
 4004 & -15 & -6 & -20 & -12 &   &   & -1 & -1 &   &   &   &   
\\ \hline
 4012 & 27 & 4 & 54 & 14 &   & 0 & 3 & 2 &   &   &   &   
\\ \hline
 4020 & -6 & 4 & -14 & 16 & 21 & 16 & -8 & -4 &   &   &   &   
\\ \hline
 4101 & -45 & -6 & -120 & -27 & -35 & -12 & -6 & -4 &   &   &   &   
\\ \hline
 5000 & 105 & 24 & 350 & 120 & 105 & 54 & 50 & 32 &   &   & 1 & 1 
\\ \hline
\end{array}
$

\end{tabular}

\end{tabular}
\bigskip

\newpage

$\Sp(3)$ ($C_3$)
\\
$
\tiny
\renewcommand{\arraycolsep}{2pt}
\begin{array}{|c||c|c||c|c||c|c||c|c||c|c||c|c||c|c||c|c|}
\hline
\lambda
 &
\multicolumn{2}{c||}{0}
 &
\multicolumn{2}{c||}{A_1}
 &
\multicolumn{2}{c||}{C_1}
 &
\multicolumn{2}{c||}{A_1 \oplus C_1}
 &
\multicolumn{2}{c||}{A_2}
 &
\multicolumn{2}{c||}{C_1 \oplus C_1}
 &
\multicolumn{2}{c||}{C_2}
 &
\multicolumn{2}{c|}{C_1 \oplus C_2}
\\ \hline
000
 & 
231 & 48 & 1232 & 384 & 462 & 144 & 924 & 432 & 511 & 360 & 275 & 120 & 81 & 60 & 54 & 54 \\ \hline
002
 & 
-18 &   & 18 & 12 & -66 & -6 & 126 & 48 & -56 & -64 & -77 & -24 & -10 & -3 &   &
\\ \hline
004
 & 
15 & 6 & 20 & 12 &   &   &   &   & 1 & 1 &   &   &   &   &   &
\\ \hline
010
 & 
-42 & -8 & -14 & -52 & -168 & -28 & -210 & -74 & 224 & -12 & -154 & -28 & -17 & -8 & -27 & -12
\\ \hline
012
 &
-24 &   & 6 & 16 & -59 & -8 & -24 & -6 &   & 8 & -12 & -2 &   &   &   &
\\ \hline
020
 & 
-36 & -4 & -84 &   & -84 & -24 & -154 & -56 & 168 & 112 & -15 & -22 & -9 & -2 & 5 & 3
\\ \hline
022
 & 
-24 & -4 & -42 & -12 & -11 & -4 & -6 & -4 & -3 & -2 &   &   &   &   &   &
\\ \hline
030
 & 
-33 & -4 & -158 & -26 & -9 & -6 & 66 & 45 & -7 & -12 & -10 & -8 & 1 &   &   &
\\ \hline
032
 & 
3 & 2 & 1 & 1 &   &   &   &   &   &   &   &   &   &   &   &
\\ \hline
040
 &
-3 & -4 & 26 &   & -18 & -12 &   & -4 & 4 & 2 & -4 & -2 &   &   &   &
\\ \hline
050
 & 
-9 & -4 & -3 & -2 & -3 & -2 & -1 & -1 &   &   &   &   &   &   &   &
\\ \hline
101
 & 
24 & 6 & -34 & 35 & 105 & 20 & 168 & 56 & -112 & 8 & 88 & 20 & -10 & -4 &   &
\\ \hline
103
 & 
9 &   & 3 & -4 & 12 & 2 & 10 & 3 &   & -2 &   &   &   &   &   &
\\ \hline
111
 & 
24 & -2 & 20 & -25 & 60 & 2 & -66 & -25 &   & -8 & 36 & 8 & 8 & 2 &   &
\\ \hline
113
 & 
-6 & -2 & -5 & -3 &   &   &   &   &   &   &   &   &   &   &   &
\\ \hline
121
 & 
27 & 2 & 54 & 7 & 27 & 4 &   & -1 & 3 & 2 & 3 &   &   &   &   &
\\ \hline
131
 & 
12 & 2 & 10 & 3 & 6 & 2 & 3 & 2 &   &   &   &   &   &   &   &
\\ \hline

200
 & 
-63 & -8 & -462 & -96 & -21 & -4 & -154 & -48 & -357 & -200 & 33 & 12 & 15 & 9 & 15 & 16
\\ \hline
202
 & 
15 & 2 & -7 & -8 & 36 & 12 & 18 & 12 & 8 & 8 & 11 & 4 &   &   &   &
\\ \hline
210
 & 
57 & 4 & 324 & 38 & 28 & 4 & 84 & 18 & 21 & 12 & -30 & -6 & 9 & 6 & -3 & -2
\\ \hline
212
 & 
3 & 2 & 13 & 9 & -3 &   & -3 & -1 &   &   &   &   &   &   &   &
\\ \hline
220
 &
 & 6 & 36 & 40 & -18 & 7 & -6 &   & -24 & -12 &   & 6 & -3 & -1 &   &
\\ \hline
222
 & 
-1 & -1 &   &   &   &   &   &   &   &   &   &   &   &   &   &
\\ \hline
230
 & 
-9 &   & -39 & -12 & 6 & 6 & -4 & -3 &   &   & 3 & 2 &   &   &   &
\\ \hline
240
 & 
3 & 2 &   &   & 1 & 1 &   &   &   &   &   &   &   &   &   &
\\ \hline
301
 & 
-54 & -4 & -144 & -10 & -72 & -14 & -18 & -5 & 7 & -2 & -28 & -12 & -6 & -4 &   &
\\ \hline
303
 & 
3 & 2 & 1 & 1 &   &   &   &   &   &   &   &   &   &   &   &
\\ \hline
311
 & 
-18 & -4 & -15 & -10 & -24 & -8 & 4 & 1 &   &   & -6 & -2 &   &   &   &
\\ \hline
321
 & 
-6 & -2 & -2 & -1 & -3 & -2 &   &   &   &   &   &   &   &   &   &
\\ \hline
400
 & 
-27 & -8 & -162 & -80 &   &   &   &   & 36 & 20 & 21 & 16 & -3 & -6 & 1 & 1
\\ \hline
402
 & 
3 &   & -3 & -4 & 4 & 2 &   &   &   &   &   &   &   &   &   &
\\ \hline
410
 & 
36 & 4 & 72 & 14 & 42 & 8 &   &   &   &   & 9 & 2 & 3 & 2 &   &
\\ \hline
420
 & 
6 & 2 & 14 & 8 &   &   &   &   &   &   & -1 & -1 &   &   &   &
\\ \hline
501
 & 3 & 2 & -6 & 1 & 7 & 4 &   &   &   &   & 3 & 2 &   &   &   &
\\ \hline
600
 & 
-21 & -8 & -28 & -16 & -21 & -12 &   &   &   &   & -5 & -4 & -1 & -1 &   &
\\ \hline
\end{array}
$
\bigskip
\\
\\

$\Spin(7)$  ($B_3$)
\\
$
\renewcommand{\arraycolsep}{2pt}
\tiny
\begin{array}{|c||c|c||c|c||c|c||c|c||c|c||c|c||c|c||c|c||c|c|}
\hline
\lambda
 & 
\multicolumn{2}{c||}{\text{\footnotesize $0$}}
 & 
\multicolumn{2}{c||}{\text{\footnotesize $A_1$}}
 & 
\multicolumn{2}{c||}{\text{\footnotesize $B_1$}}
 & 
\multicolumn{2}{c||}{\text{\footnotesize $A_1 \oplus B_1$}}
 & 
\multicolumn{2}{c||}{\text{\footnotesize $A_2$}}
 & 
\multicolumn{2}{c||}{\text{\footnotesize $B_2$}}
 & 
\multicolumn{2}{c||}{\text{\footnotesize $D_2$}}
 & 
\multicolumn{2}{c||}{\text{\footnotesize $D_2 \oplus B_1$}}
 & 
\multicolumn{2}{c|}{\text{\footnotesize $D_3$}}
\\ \hline
000
 & 
231 & 48 & 1001 & 312 & 693 & 216 & 1155 & 540 & 246 & 160 & 126 & 108 & 330 & 144 & 470 & 312 & 7 & 8
\\ \hline
002
 & 
-27 & 6 & -99 & 36 & -63 & 24 & 231 & 144 & -25 & 5 & -60 & -36 & 21 & 33 & 214 & 144 & -1 & -1
\\ \hline
004
 & 
-24 &   & 1 & 24 & -162 & -48 & -60 & -36 & -1 & 4 &   &   &   &   & 20 & 12 &   & 
\\ \hline
006
 & 
15 & 6 & 20 & 12 &   &   &   &   & 1 & 1 &   &   &   &   &   &   &   &
\\ \hline
010
 & 
-63 & -8 & -147 & -40 & -336 & -52 & -616 & -120 & 51 &   & -18 & -24 & -63 & -14 & -327 & -69 & 3 & 2
\\ \hline
012
 & 
-3 & -4 & -39 & -22 & 56 & -8 &   & -12 & -6 & -10 & 10 & 3 & 14 & 4 & -45 & -12 &   &
\\ \hline
014
 & 
-27 & -4 & -45 & -12 & -9 & -2 &   &   & -3 & -2 &   &   &   &   &   &   &   &
\\ \hline
020
 & 
-36 & -8 & -108 & -36 & -60 & -36 & -12 & -34 & 12 & 8 & 3 & 6 & -42 & -8 & 79 & 36 & &   \\ \hline
022
 & 
15 &   & 36 & 4 & -7 & -8 & -21 & -16 & 3 & 1 &   &   & -5 & -3 &   &   &   &
\\ \hline
030
 & 
-15 & -4 & -23 & -8 & -24 & -18 & 24 &   & -1 &   &   &   & 15 & 6 & 1 & 1 &   &
\\ \hline
032
 & 
3 & 2 & 1 & 1 &   &   &   &   &   &   &   &   &   &   &   &   &   &
\\ \hline
040
 & 
-9 & -4 & -3 & -2 & -3 & -2 & -1 & -1 &   &   &   &   &   &   &   &   &   &
\\ \hline
100
 & 
-21 & -8 & -175 & -60 & 126 & -16 &   & -72 & -82 & -40 & 60 & 14 & -126 & -40 & -120 & -72 & -5 & -4
\\ \hline
102
 & 
57 & 4 & 123 & 12 & 229 & 34 & 189 & 48 & -3 & -2 & -4 &   & -28 & -12 & -27 & -14 &   &   \\ \hline
104
 & 
18 & 2 & 24 & 5 & 42 & 8 & 35 & 12 & 3 & 2 &   &   &   &   &   &   &   &
\\ \hline
110
 & 
36 & 4 & 198 & 30 & -84 &   & -168 & -32 & 35 & 20 & 15 & 11 & 84 & 8 & -10 & -32 &   &   
\\ \hline
112
 & 
18 & 2 & 27 & 2 & 48 & 17 & -12 & 8 & 1 & 2 &   &   & 6 & 2 & -5 & -3 &   &
\\ \hline
114
 & 
-6 & -2 & -5 & -3 &   &   &   &   &   &   &   &   &   &   &   &   &   &
\\ \hline
120
 & 
-9 & 4 & -21 & 16 & -12 & 12 & 56 & 32 & -9 & -4 & -3 & -1 & -18 & -4 & 6 &   &   &
\\ \hline
122
 & 
12 & 2 & 12 & 4 & 4 & 1 & 3 & 2 &   &   &   &   &   &   &   &   &   &
\\ \hline
130
 & -6 &   & -11 & -2 & 6 & 4 &   & 2 &   &   &   &   &   &   &   &   &   &
\\ \hline
200
 & 
-42 &   & -210 & -16 & 28 & 32 & 84 & 88 & -90 & -48 & 28 & 32 & 13 & 16 & 144 & 112 &   &
\\ \hline
202
 & 
-48 & -4 & -118 & -20 & -84 & -8 & -84 & -32 & -4 & -3 & -6 & -4 & -12 & -7 & 22 & 16 & &
\\ \hline
204
 & 
9 & 4 & 12 & 10 & 3 & 2 & -5 & -3 &   &   &   &   &   &   &   &   &   &
\\ \hline
210
 & 
27 & 2 & 117 & 20 & -45 & -15 & 3 & -20 & 6 &   &   & -4 & 36 & 16 & -30 & -11 & &
\\ \hline
212
 & 
-15 & -2 & -17 & -5 & -21 & -4 &   &   &   &   &   &   & 3 & 2 &   &   &   &
\\ \hline
220
 & 
 &   & -9 & -4 & 18 & 8 & -12 & -10 &   &   &   &   & -9 & -4 &   &   &   &
\\ \hline
222
 & 
-1 & -1 &   &   &   &   &   &   &   &   &   &   &   &   &   &   &   &
\\ \hline
230
 & 
3 & 2 &   &   & 1 & 1 &   &   &   &   &   &   &   &   &   &   &   &
\\ \hline
300
 &
-18 & -8 & -90 & -52 & 42 &   & 126 & 72 & 14 & 8 & -3 & -9 & -63 & -36 & -9 &   &   &   
\\ \hline
302
 & 
-9 & -4 & -9 & -8 & -24 & -18 & 24 & 12 &   &   &   &   & -6 & -6 & 1 & 1 &   &
\\ \hline
304
 & 
3 & 2 & 1 & 1 &   &   &   &   &   &   &   &   &   &   &   &   &   &
\\ \hline
310
 & 
36 & 4 & 66 & 16 & 48 & 4 & -14 & -2 &   &   & 3 & 2 & 18 & 8 & -3 & -2 &   &
\\ \hline
312
 & 
-6 & -2 & -3 & -2 & -2 & -1 &   &   &   &   &   &   &   &   &   &   &   &
\\ \hline
320
 & 
3 &   & 4 & 2 & -3 & -4 &   &   &   &   &   &   &   &   &   &   &   &
\\ \hline
400
 & 
-9 &   & -45 & -16 & 39 & 32 & -15 & -8 &   &   &   & 2 & -9 & -8 & 5 & 4 &   &
\\ \hline
402
 & 
6 & 2 &   &   & 14 & 8 &   &   &   &   &   &   & -1 & -1 &   &   &   &
\\ \hline
410
 & 
3 & 2 & 7 & 4 & -6 & 1 &   &   &   &   &   &   & 3 & 2 &   &   &   &
\\ \hline
500
 & 
-21 & -8 & -21 & -12 & -28 & -16 &   &   &   &   & -1 & -1 & -5 & -4 &   &   &   &
\\ \hline
\end{array}
$

$\Spin(8)$ ($D_4$)
\\
$
\tiny
\renewcommand{\arraycolsep}{2pt}
\begin{array}{|c||c|c||c|c||c|c||c|c||c|c||c|c||c|c||c|c||c|c|}
\hline
\lambda
 & 
\multicolumn{2}{c||}{0}
 & 
\multicolumn{2}{c||}{A_1}
 & 
\multicolumn{2}{c||}{A_1 \oplus A_1}
 & 
\multicolumn{2}{c||}{A_2}
 & 
\multicolumn{2}{c||}{D_2}
 & 
\multicolumn{2}{c||}{A_1 \oplus D_2}
 & 
\multicolumn{2}{c||}{A_3}
 & 
\multicolumn{2}{c||}{D_3}
 & 
\multicolumn{2}{c|}{D_2 \oplus D_2}
\\ \hline
 0000 & 1617 & 192 & 15092 & 2688 & 6468 & 1728 & 8547 & 3360 & 6468 & 1728 & 12936 & 5184 & 381 & 352 & 381 & 352 & 3430 & 1944 
\\ \hline
 0002 & -189 & 24 & -1512 & 312 & 1134 & 438 & -1749 & -48 &   & 288 & 3234 & 1584 & 111 & 105 & -135 & -96 & 1470 & 864 
\\ \hline
 0004 & 27 & 24 & 162 & 240 & 483 & 240 & -30 & 132 & -462 & -144 & -154 & -144 & 3 & 12 &   &   & 105 & 54 
\\ \hline
 0006 & 105 & 24 & 350 & 120 & 105 & 54 & 50 & 32 &   &   &   &   & 1 & 1 &   &   &   &   
\\ \hline
 0020 & -189 & 24 & -1512 & 312 &   & 288 & -1749 & -48 &   & 288 & 3234 & 1584 & -135 & -96 & -135 & -96 & 1470 & 864 
\\ \hline
 0022 & -324 & -12 & -2052 & -132 & -420 & -48 & -108 & 40 & -996 & -183 & -336 & -120 & -10 & -8 & 11 & 4 & 245 & 144 
\\ \hline
 0024 & 54 & 12 & 216 & 84 & 105 & 72 & 9 & 12 & -84 & -24 & -70 & -36 &   &   &   &   &   &   
\\ \hline
 0040 & 27 & 24 & 162 & 240 & -462 & -144 & -30 & 132 & -462 & -144 & -154 & -144 &   &   &   &   & 105 & 54 
\\ \hline
 0042 & 54 & 12 & 216 & 84 & -30 & -9 & 9 & 12 & -84 & -24 & -70 & -36 &   &   &   &   &   &   
\\ \hline
 0044 & -15 & -6 & -20 & -12 &   &   & -1 & -1 &   &   &   &   &   &   &   &   &   &   
\\ \hline
 0060 & 105 & 24 & 350 & 120 &   &   & 50 & 32 &   &   &   &   &   &   &   &   &   &   
\\ \hline
 0100 & -588 & -32 & -4900 & -432 & -2940 & -296 & -462 & -408 & -2940 & -296 & -8526 & -988 & 38 & -24 & 38 & -24 & -3136 & -432 
\\ \hline
 0102 & 27 & -8 & 180 & -92 & -420 & -68 & -207 & -136 & 630 & 12 & 84 & -6 & -9 & -18 & 15 & 4 & -441 & -72 
\\ \hline
 0104 & -162 & -8 & -702 & -60 & -273 & -44 & -108 & -24 & 42 & 4 & 84 & 14 & -3 & -2 &   &   &   &   
\\ \hline
 0120 & 27 & -8 & 180 & -92 & 630 & 12 & -207 & -136 & 630 & 12 & 84 & -6 & 15 & 4 & 15 & 4 & -441 & -72 
\\ \hline
 0122 & -108 & -4 & -504 & -34 & -90 & -24 & -90 & -24 & 78 & -4 & 186 & 30 &   &   &   &   &   &   
\\ \hline
 0124 & 27 & 4 & 54 & 14 &   &   & 3 & 2 &   &   &   &   &   &   &   &   &   &   
\\ \hline
 0140 & -162 & -8 & -702 & -60 & 42 & 4 & -108 & -24 & 42 & 4 & 84 & 14 &   &   &   &   &   &   
\\ \hline
 0142 & 27 & 4 & 54 & 14 &   &   & 3 & 2 &   &   &   &   &   &   &   &   &   &   
\\ \hline
 0200 & -216 & -24 & -1512 & -288 & -420 & -64 & 18 & -312 & -420 & -64 & 490 & 256 & 30 & 24 & 30 & 24 & 1029 & 364 
\\ \hline
 0202 & 54 & -4 & 270 & -36 & 225 & 47 & 132 & -32 & -180 & -64 & -384 & -128 & 3 & 1 &   &   & 21 & 16 
\\ \hline
 0204 & -6 & 4 & -14 & 16 & 21 & 16 & -8 & -4 &   &   &   &   &   &   &   &   &   &   
\\ \hline
 0220 & 54 & -4 & 270 & -36 & -180 & -64 & 132 & -32 & -180 & -64 & -384 & -128 &   &   &   &   & 21 & 16 
\\ \hline
 0222 & -54 & -4 & -144 & -20 & 9 & 8 & -9 & -4 & -6 & -1 & -6 & -4 &   &   &   &   &   &  
\\ \hline
 0240 & -6 & 4 & -14 & 16 &   &   & -8 & -4 &   &   &   &   &   &   &   &   &   &   
\\ \hline
 0300 & -87 & -24 & -464 & -228 & -30 & -90 & 177 & 152 & -30 & -90 & 570 & 27 & -1 &   & -1 &   & -49 &   
\\ \hline
 0302 & 27 &   & 81 &   & -57 & -18 & 24 & 8 & -30 & -18 & -4 & -9 &   &   &   &   &   &   
\\ \hline
 0320 & 27 &   & 81 &   & -30 & -18 & 24 & 8 & -30 & -18 & -4 & -9 &   &   &   &   &   &   
\\ \hline
 0322 & -3 & -2 & -1 & -1 &   &   &   &   &   &   &   &   &   &   &   &   &   &   
\\ \hline
 0400 & -51 & -8 & -170 & -48 & 75 & 8 & -48 & -12 & 75 & 8 & 6 & -8 &   &   &   &   & 1 & 1 
\\ \hline
 0402 & 9 & 4 & 6 & 4 & 1 & 1 &   &   &   &   &   &   &   &   &   &   &   &   
\\ \hline
 0420 & 9 & 4 & 6 & 4 &   &   &   &   &   &   &   &   &   &   &   &   &   &   
\\ \hline
 0500 & -27 & -8 & -27 & -12 & -3 & -2 &   &   & -3 & -2 & -1 & -1 &   &   &   &   &   &   
\\ \hline
 1011 & 606 & 10 & 4444 & 125 & 798 & 4 & 1760 & 244 & 798 & 4 & -294 & -248 & -2 & 12 & -2 & 12 & -392 & -192 
\\ \hline
 1013 & 162 &   & 864 &   & 75 & -26 & 189 & 18 & 300 & 46 & 150 & 49 & 6 & 4 &   &   & -35 & -12 
\\ \hline
 1015 & -45 & -6 & -120 & -27 & -35 & -12 & -6 & -4 &   &   &   &   &   &   &   &   &   &  
\\ \hline
 1031 & 162 &   & 864 &   & 300 & 46 & 189 & 18 & 300 & 46 & 150 & 49 &   &   &   &   & -35 & -12 
\\ \hline
 1033 & -27 & -4 & -81 & -22 & -15 & -6 & -6 & -6 & 12 & 2 & 10 & 3 &   &   &   &   &   &  
\\ \hline
 1051 & -45 & -6 & -120 & -27 &   &   & -6 & -4 &   &   &   &   &   &   &   &   &   &   
\\ \hline
 1111 & 96 & 14 & 560 & 147 & 120 & 60 & -320 & 24 & 120 & 60 & 60 & 63 & -8 & -2 & -8 & -2 &   & -32 
\\ \hline
 1113 & 54 &   & 189 &   & 36 & -10 & 24 & 8 & 36 & 8 & 18 & 11 &   &   &   &   &   &   
\\ \hline
 1131 & 54 &   & 189 &   & 36 & 8 & 24 & 8 & 36 & 8 & 18 & 11 &   &   &   &   &   &   
\\ \hline
 1133 & 6 & 2 & 5 & 3 &   &   &   &   &   &   &   &   &   &   &   &   &   &   
\\ \hline
 1211 & 27 & 6 & 108 & 45 & 21 & 20 & -45 & -6 & 21 & 20 & 16 & 25 &   &   &   &   & -3 & -2 
\\ \hline
 1213 & -12 & -2 & -16 & -5 & -3 & -2 &   &   &   &   &   &   &   &   &   &   &   &   
\\ \hline
 1231 & -12 & -2 & -16 & -5 &   &   &   &   &   &   &   &   &   &   &   &   &   &   
\\ \hline
 1311 & 24 & 2 & 44 & 7 & 6 & 2 &   &   & 6 & 2 & 3 & 2 &   &   &   &   &   &   
\\ \hline
 2000 & -189 & 24 & -1512 & 312 &   & 288 & -1749 & -48 & 1134 & 438 & 3234 & 1584 & -135 & -96 & 111 & 105 & 1470 & 864 
\\ \hline
 2002 & -324 & -12 & -2052 & -132 & -420 & -48 & -108 & 40 & -420 & -48 & -336 & -120 & -10 & -8 & -10 & -8 & 245 & 144 
\\ \hline
 2004 & 54 & 12 & 216 & 84 & 105 & 72 & 9 & 12 & -30 & -9 & -70 & -36 &   &   &   &   &   &   
\\ \hline
 2020 & -324 & -12 & -2052 & -132 & -996 & -183 & -108 & 40 & -420 & -48 & -336 & -120 & 11 & 4 & -10 & -8 & 245 & 144 
\\ \hline
 2022 & -69 & -2 & -299 & -16 & -111 & -8 & 14 & 20 & -111 & -8 & -186 & -80 &   &   &   &   & 5 & 3 
\\ \hline
 2024 & -9 & -4 & -15 & -12 & 5 & 3 &   &   &   &   &   &   &   &   &   &   &   &   
\\ \hline
 2040 & 54 & 12 & 216 & 84 & -84 & -24 & 9 & 12 & -30 & -9 & -70 & -36 &   &   &   &   &   &   
\\ \hline
 2042 & -9 & -4 & -15 & -12 &   &   &   &   &   &   &   &   &   &   &   &   &   &   
\\ \hline
 2100 & 27 & -8 & 180 & -92 & 630 & 12 & -207 & -136 & -420 & -68 & 84 & -6 & 15 & 4 & -9 & -18 & -441 & -72 
\\ \hline
 2102 & -108 & -4 & -504 & -34 & -90 & -24 & -90 & -24 & -90 & -24 & 186 & 30 &   &   &   &   &   &   
\\ \hline
 2104 & 27 & 4 & 54 & 14 &   &   & 3 & 2 &   &   &   &   &   &   &   &   &   &   
\\ \hline
 2120 & -108 & -4 & -504 & -34 & 78 & -4 & -90 & -24 & -90 & -24 & 186 & 30 &   &   &   &   &   &   
\\ \hline
 2122 & 33 & 6 & 77 & 27 & -6 & -2 &   &   & -6 & -2 & -3 & -1 &   &   &   &   &   &   
\\ \hline
 2140 & 27 & 4 & 54 & 14 &   &   & 3 & 2 &   &   &   &   &   &   &   &   &   &   
\\ \hline
 2200 & 54 & -4 & 270 & -36 & -180 & -64 & 132 & -32 & 225 & 47 & -384 & -128 &   &   & 3 & 1 & 21 & 16 
\\ \hline
 2202 & -54 & -4 & -144 & -20 & 9 & 8 & -9 & -4 & 9 & 8 & -6 & -4 &   &   &   &   &   &   
\\ \hline
 2220 & -54 & -4 & -144 & -20 & -6 & -1 & -9 & -4 & 9 & 8 & -6 & -4 &   &   &   &   &   &  
\\ \hline
 2222 & 1 & 1 &   &   &   &   &   &   &   &   &   &   &   &   &   &   &   &   
\\ \hline
 2300 & 27 &   & 81 &   & -30 & -18 & 24 & 8 & -57 & -18 & -4 & -9 &   &   &   &   &   &   
\\ \hline
 2302 & -3 & -2 & -1 & -1 &   &   &   &   &   &   &   &   &   &   &   &   &   &   
\\ \hline
 2320 & -3 & -2 & -1 & -1 &   &   &   &   &   &   &   &   &   &   &   &   &   &   
\\ \hline
 2400 & 9 & 4 & 6 & 4 &   &   &   &   & 1 & 1 &   &   &   &   &   &   &   &   
\\ \hline
 3011 & 162 &   & 864 &   & 300 & 46 & 189 & 18 & 75 & -26 & 150 & 49 &   &   & 6 & 4 & -35 & -12 
\\ \hline
 3013 & -27 & -4 & -81 & -22 & -15 & -6 & -6 & -6 & -15 & -6 & 10 & 3 &   &   &   &   &   &   
\\ \hline
 3031 & -27 & -4 & -81 & -22 & 12 & 2 & -6 & -6 & -15 & -6 & 10 & 3 &   &   &   &   &   &  
\\ \hline
 3033 & -3 & -2 & -1 & -1 &   &   &   &   &   &   &   &   &   &   &   &   &   &   
\\ \hline
 3111 & 54 &   & 189 &   & 36 & 8 & 24 & 8 & 36 & -10 & 18 & 11 &   &   &   &   &   &   
\\ \hline
 3113 & 6 & 2 & 5 & 3 &   &   &   &   &   &   &   &   &   &   &   &   &   &   
\\ \hline
 3131 & 6 & 2 & 5 & 3 &   &   &   &   &   &   &   &   &   &   &   &   &   &   
\\ \hline
 3211 & -12 & -2 & -16 & -5 &   &   &   &   & -3 & -2 &   &   &   &   &   &   &   &   
\\ \hline
 4000 & 27 & 24 & 162 & 240 & -462 & -144 & -30 & 132 & 483 & 240 & -154 & -144 &   &   & 3 & 12 & 105 & 54 
\\ \hline
 4002 & 54 & 12 & 216 & 84 & -30 & -9 & 9 & 12 & 105 & 72 & -70 & -36 &   &   &   &   &   &   
\\ \hline
 4004 & -15 & -6 & -20 & -12 &   &   & -1 & -1 &   &   &   &   &   &   &   &   &   &   
\\ \hline
 4020 & 54 & 12 & 216 & 84 & -84 & -24 & 9 & 12 & 105 & 72 & -70 & -36 &   &   &   &   &   &   
\\ \hline
 4022 & -9 & -4 & -15 & -12 &   &   &   &   & 5 & 3 &   &   &   &   &   &   &   &   
\\ \hline
\end{array}
$

\newpage

$\Spin(8)$ ($D_4$) (continued)
\\
$
\tiny
\renewcommand{\arraycolsep}{2pt}
\begin{array}{|c||c|c||c|c||c|c||c|c||c|c||c|c||c|c||c|c||c|c|}
\hline
\lambda
 & 
\multicolumn{2}{c||}{0}
 & 
\multicolumn{2}{c||}{A_1}
 & 
\multicolumn{2}{c||}{A_1 \oplus A_1}
 & 
\multicolumn{2}{c||}{A_2}
 & 
\multicolumn{2}{c||}{D_2}
 & 
\multicolumn{2}{c||}{A_1 \oplus D_2}
 & 
\multicolumn{2}{c||}{A_3}
 & 
\multicolumn{2}{c||}{D_3}
 & 
\multicolumn{2}{c|}{D_2 \oplus D_2}
\\ \hline
 4040 & -15 & \hspace{0.5em}-6\hspace{0.5em} & -20 & -12 & \phantom{500} &
 \phantom{500} & \hspace{0.5em}-1\hspace{0.5em} & \hspace{0.5em}-1\hspace{0.5em}
 & 
 \phantom{500} & \phantom{500} & \phantom{500} & \phantom{500} & \phantom{500} &
 \phantom{500} & \phantom{500} & \phantom{500} & \phantom{500} & \phantom{500}
\\ \hline
 4100 & -162 & -8 & -702 & -60 & 42 & 4 & -108 & -24 & -273 & -44 & 84 & 14 &   &   & -3 & -2 &   &   
\\ \hline
 4102 & 27 & 4 & 54 & 14 &   &   & 3 & 2 &   &   &   &   &   &   &   &   &   &   
\\ \hline
 4120 & 27 & 4 & 54 & 14 &   &   & 3 & 2 &   &   &   &   &   &   &   &   &   &   
\\ \hline
 4200 & -6 & 4 & -14 & 16 &   &   & -8 & -4 & 21 & 16 &   &   &   &   &   &   &   &   
\\ \hline
 5011 & -45 & -6 & -120 & -27 &   &   & -6 & -4 & -35 & -12 &   &   &   &   &   &   &   &  
\\ \hline
 6000 & 105 & 24 & 350 & 120 &   &   & 50 & 32 & 105 & 54 &   &   &   &   & 1 & 1 &   &   
\\ \hline
\end{array}
$

\end{appendix}

\end{document}